\documentclass{article}

\usepackage[a4paper, margin=0.8in]{geometry}
\usepackage[utf8]{inputenc}
\usepackage[dvipsnames,table,xcdraw]{xcolor}
\usepackage[affil-it]{authblk}
\usepackage[misc]{ifsym}
\usepackage{sectsty,verbatim,listings,braket,enumitem, kantlipsum,graphicx,multicol,mathtools,csquotes,amsmath,mathtools,url,subfiles,qcircuit,amsthm,amssymb,fdsymbol,rotating,caption,hyperref,fontawesome,xcolor,arydshln,hyperref,appendix,xfrac,todonotes,caption,subcaption}

\definecolor{darkblue}{rgb}{0.0, 0.0, 0.55}

\def\titlefont{\color{darkblue}}
\sectionfont{\color{darkblue}}
\subsectionfont{\color{darkblue}}
\subsubsectionfont{\color{darkblue}}

\let\OLDthebibliography\thebibliography
\renewcommand\thebibliography[1]{
  \OLDthebibliography{#1}
  \setlength{\parskip}{0pt}
}

\title{\titlefont \textbf{\Large Visualizing Quantum Circuit Probability}
\\ \Large estimating computational action for quantum program synthesis}

\usepackage[affil-it]{authblk}

\author[1,6]{Bao Gia Bach}
\author[2,3,6]{Akash Kundu}
\author[4,6]{Tamal Acharya}
\author[5,6, \Letter]{Aritra Sarkar}
\affil[1]{Faculty of Computer Science and Engineering, Ho Chi Minh City University of Technology, Viet Nam}
\affil[2]{Joint Doctoral School, Silesian University of Technology, Gliwice, Poland}
\affil[3]{Institute of Theoretical and Applied Informatics, Polish Academy of Sciences, Gliwice, Poland}
\affil[4]{Independent Researcher, Bengaluru, India}
\affil[5]{Quantum Machine Learning research group, Quantum Computing division, QuTech, The Netherlands}
\affil[6]{Quantum Intelligence research team,  
Department of Quantum \& Computer Engineering, \protect\\Delft University of Technology, The Netherlands}

\affil[ \Letter ]{a.sarkar-3@tudelft.nl}
\date{} %\today}

\begin{document}

\maketitle

\begin{abstract}

This research applies concepts from algorithmic probability to Boolean and quantum combinatorial logic circuits. 
A tutorial-style introduction to states and various notions of the complexity of states are presented.
Thereafter, the probability of states in the circuit model of computation is defined.
Classical and quantum gate sets are compared to select some characteristic sets.
The reachability and expressibility in a space-time-bounded setting for these gate sets are enumerated and visualized.
These results are studied in terms of computational resources, universality and quantum behavior.
The article suggests how applications like geometric quantum machine learning, novel quantum algorithm synthesis and quantum artificial general intelligence can benefit by studying circuit probabilities. 

\end{abstract}


\section{Introduction}

% motivate why others should bother, a problems others can appreciate, instead of a dry theoretical work... e.g. NISQ problems of decoherence (cite: Preskill, Leymann)

Quantum computing has entered a technological readiness level where quantum processor platforms, albeit limited, are becoming accessible for experimentation.
This rapid progress has encouraged researchers to study various real-world industrial/scientific applications~\cite{bertels2021quantum} using quantum algorithms.
The logical formulations of these algorithms are then processed by a quantum computing stack~\cite{bertels2020quantum} of system abstractions into low-level quantum circuits for a gate-based quantum computing device.
The so-called NISQ (noisy intermediate-scale quantum) era~\cite{preskill2018quantum,leymann2020bitter} characterizes the limitations of current quantum processors in coherence time, gate errors, and qubit connectivity.
This has led to explorations from the other end~\cite{shi2020resource}, in devising design strategies and finding use cases for these limited computing power to achieve a computational advantage.
To better utilize these limited devices, it is imperative to understand the relations between quantum logic and physical resources.
This motivates the research presented in this article.

Quantum computation lies at the intersection of quantum physics and computer science.
It has allowed a rich exchange of concepts between these two fields.
Specific to the interests of this research, (i) physical laws provide fundamental bounds to computation, while (ii) computation provides an information-theoretic explanation of many physical phenomena.
The former was first explored in the context of thermodynamic limits~\cite{landauer1961irreversibility}, leading to the development of reversible computation~\cite{bennett1973logical,fredkin1982conservative}, and eventually to define the limits of quantum computation~\cite{lloyd2000ultimate,markov2014limits}.
Efforts of the latter come under the purview of digital physics.
Some seminal contributions include cellular automaton~\cite{wolfram2002new} and constructor theory~\cite{deutsch2013constructor}, informational axioms~\cite{hardy2001quantum,muller2020law} of quantum mechanics, a principle of stationary action for computing~\cite{toffoli2018action,lloyd2016universal}, and tensor networks, among others.
The work presented in this research is foundational and would have applications for both directions of this synergy.
Our focus is on an empirical demonstration of the consequences of existing theoretical ideas for quantum computing.
% not on rigorous theoretical derivation, rather, w
Specifically, we transport concepts from algorithmic information theory~\cite{solomonoff1960preliminary} (a sub-field of theoretical computer science and artificial intelligence~\cite{hutter2004universal}) to gate-based quantum computation~\cite{deutsch1989quantum}.

% universal gates are easy to engineer, or atleast finite number of gates are only possible, means we can only do certain algorithms.

In this work, we consider an enumeration of the space of quantum circuits (as QASM codes).
This is a very small sample of the uncountable infinite quantum processes that can be defined on the Hilbert space of a given dimension. 
The subset of processes is based on (i) the native gate set (of a quantum processor), (ii) the maximum circuit width (total number of qubits) and, (iii) the bounds on the circuit depth (based on the decoherence time).
We investigate how these circuits map a space of classical inputs to classical outputs (via Z-axis measurements).
Our formalism assumes the QASM/circuits are encoded using a discrete universal gate set.
The space of quantum programs constructed in such a matter is enumerable and is thus formally countably infinite.
However, in most practical implementations (with finite and circuit depth), the set of meaningful computations is finite.
Even for universal gate sets with arbitrary rotation angles, a finite number of control signal configurations in practical quantum computer implementation effectively discretizes the set of native gates.
While any arbitrary unitary can be decomposed to arbitrary precision given a universal gate set (e.g., $\{Rx(\theta_x), Ry(\theta_y), CX\}$), given resource bounds (e.g., in lines of QASM codes before the system decoheres) the space of programs cannot map to any quantum process.
This limits us from exploring functional physical processes that can be efficiently described in terms of quantum gates to be simulated on a quantum computer.

% (so what is our value proposition?) can a put our understanding of `what's possible and what's not' in a more rigorous footing?

Since the first quantum algorithms were formulated in the 1990s, the discovery of new algorithms~\cite{jordan} has progressed steadily.  %(e.g. Deutsch, Jozsa, Simon, Shor, Grover) (e.g. QuantumAlgorithmZoo).
However, quantum algorithm design involves quantum mechanical phenomena (e.g., superposition, entanglement), which is counter-intuitive to human experience.
Thus, reasoning in terms of mathematical formalism has been a barrier to entry to develop more advanced quantum logic and is thus a bottleneck for broader adoption of quantum accelerated computing.
There have been some proposals to remedy these issues via genetic programming~\cite{spector2004automatic} and circuit synthesis~\cite{quetschlich2022towards}.
In this work, we carry forward this research direction via the principled approach~\cite{sarkar2022automated} of algorithmic information theory.
A quantum program synthesis framework would require understanding the space of quantum programs and their associated resources, to implement the program search/induction.
As discussed in this article, the landscape of resource-bounded quantum circuits and their corresponding classical information processing capability, lay the groundwork towards this.

The rest of the article is organized as follows.
In \S~\ref{s:states}, we describe how states are represented as symbols and transformations over these symbols, and how this affects their statistical and algorithmic complexities. 
\S~\ref{s:circuits} discusses the subtleties of forming Boolean and quantum circuits from gate sets. 
In \S~\ref{s:implementation}, we present our implementation of the enumeration of state complexities using various gate sets.
The results are visualized and analyzed. 
\S~\ref{s:application} concludes the article with a discussion of various applications of this research. 
% \S~\ref{s:conclusion} concludes the article. 

\section{States and complexities} \label{s:states}

% We start with a brief discussion of statistical and algorithmic complexity/entropy.
The states of a system define its observable behavior.
These can be encoded in various ways.
A common way to encode them is by assigning a symbol for each unique/distinguishable state.
Subsequent observations of the same system, (or a larger system composed of systems of this size) are denoted by a string of this alphabet.
A system with a single state is not very interesting, nothing really happens there.
A very simple case of a slightly more interesting system is a coin, with two states - heads and tails.
Coins can be fair or biased, can be tossed multiple times, or multiple coins can be tossed together/consecutively/conditionally.
The outcome of a series of (or in parallel) coin tosses can be represented by a Boolean string.
Given an ensemble of all Boolean strings of length $n$, represented as $\{0,1\}^{\otimes n}$, each string might be equally probable with $\sfrac{1}{2^n}$.
In a physical system, if each output is equally likely, the uniform distribution models that system, e.g. a communication line might need to transfer each encoding with equal probability.
According to this, a fair coin tossed 8 times, would have the same probability and element of surprise for $1111111011$ as that of a specific permutation of 5 $1$s and 5 $0$s, e.g. $0101100101$.

If everything is equally likely and unrelated to each other, it is a very boring grey world.
Thankfully it is not so.
We perceive structures around us.
Why that is this way is hard to answer - but a likely explanation is that our human biological/technological sensing tools are limited.
So instead of parsing the underlying randomness in its full spectrum, we perceive an emergent statistical structure.
These structures allow us two additional ways of enhancing our representation of states.
The complexity of states can be studied from these two perspectives - statistical and algorithmic.

\subsection{The statistical emergence of entropy}

The first enhancement is based on relaxing the criteria of all states being `equally likely'.
We find that we have apparent favoritism towards $0101100101$ being a more acceptable result of a series of fair coin tosses.
This is based on our ignorance towards the micro-states of the permutations.
We focus on the pattern that the total possible states with 9 $1$s are $10$, while those with 5 $1$s and 5 $0$s are $10*9*8*7$, similar to entropy in statistical thermodynamics.
States with higher entropy are more in number and this flow towards an expected higher entropy state in the universe is what gives us our perception of time.
In information theory, given a discrete random variable $X$, which takes values in the alphabet $\mathcal{X}$ and is distributed according to $p:\mathcal{X} \rightarrow [0,1]$, the Shannon entropy~\cite{shannon1948mathematical} of the variable sampled from this ensemble is given by $H(X) = -\sum_{x \in \mathcal{X}} p(x)\log p(x)$. 
This likeness denotes the average level of statistical information, or the surprise/uncertainty inherent to the variable's possible outcomes, and is the maximum for a uniform distribution.
The way to optimally encode a biased set of percepts as information is the basis of code words like Huffman coding.
The encoding is designed to tune to the bit lengths of concepts by making the most used concepts more economical.
To balance it, less used concepts become more costly than their native length.
E.g. the probabilities $p(00)=0.4, p(01)=0.05, p(10)=0.2, p(11)=0.35$ is best encoded as $00:\rightarrow 0, 01:\rightarrow 111, 10:\rightarrow 110, 11:\rightarrow 10$.
Note that, this code word is better than the original only as long as the biased probability distribution is maintained (which in turn might be an artifact of sensing emergent macro-states).
If instead all strings are equally probable, these coding schemes are more costly.

We do something similar with semantics in languages, e.g. instead of having new words for every single huge animal with a large truck with a specific set of $x,y,z,..$ features that we fail to distinguish, we call all of them `an elephant', while that we can distinguish, e.g. your pet cat, we give them special names.
Your friend might not be able to distinguish your cat from another, or to the eyes of a trained mahout, every elephant is uniquely identifiable - leading to the subjectivity of language.
Similarly, using the scientific names of species or the full genetic code of an individual, would be tedious for everyday use, however is useful for biological classification or medical treatment.
Thus, much the same way as Huffman coding, words in languages arise due to do different ways of ignoring details and focusing on specific emergent semantics.
% We will return to this subjectivity in universal Turing machines, compilers and gate sets, later.
The compression provided by a language forms the basis of comprehension, much the same way macro states (micro states preserving certain symmetries/features) leads to emergent physical laws.

\subsection{The algorithmic emergence of universality}

The second enhancement is based on relaxing the criteria of all states being `unrelated'.
While an unique encoding for all percepts under consideration is good, we can compress and comprehend better if we can find relation among these symbols.
For example, we can relate bit strings by their inverses, or arrange integers consecutively on a number line.
Assigning symbols to relations is merely an attempt to minimize the number of code words itself by ignoring symbols for states that can now be described by some specific syntactic composite of symbols of other states and relations.
To map to every percept, the total length of the encodings using these codes is not necessarily less than the original binary or Huffman coding.
Thus, again, it is subjective when these relations will be beneficial instead of adding extra complexity to the encoding.
The goal is not about being the most resource-efficient way for the full spectrum of percepts, but rather using a smaller set of symbols for a biased distribution or subset of percepts.
This trade-off between generality and efficiency is the reason esoteric languages like MetaGolfScript~\cite{MetaGolfScript} are banned from code golf contests, or the RISC and CISC architectures exist in tandem.

However, surprisingly, we find that, some relations are so ubiquitous that, it can map to all percepts (often even from an infinite set) with just the encoding of the relation and that of a starting percept.
For example, the successor function (i.e. add 1) is represented by $1$ and the number 0 is represented by $0$. 
With this, any natural number can be represented by nesting the successor function, e.g., $1110$ is 3.
While the successor function is universal over the set of natural numbers given 0, the multiplication operation with the set of all prime numbers can also span the natural numbers.
Such set of inputs and transformations are called universal for the target output set.
Some of these symbols (of states and relations) are so powerful that they can potentially represent an infinite set of states with a finite set of symbols and some compositional rules, e.g. any d-base numeral system with d-symbols and a positional notation can represent any integer.

From an engineering point of view, having such universal set of states and transformations helps in taming an infinitude of possibility with a finite number of these building blocks, e.g., a keyboard is made of alphabets instead of a key for every word in the dictionary.
There are two subtleties to this enhancement.
(i) Firstly, since we now have a finite number of block types to construct an infinite number of states, the length of the description using these blocks can potentially be infinite.
If we put a bound on the number of blocks we can use, by combinatorial arguments, it bounds the number of states we can describe.
States that require longer descriptions are not expressible.
(ii) Secondly, while the original states represented an observable behavior, these new pseudo-states and transformations that we introduced to reduce our symbol set need not necessarily have intuitive standalone meaning. 
For some, it may, for example, the bit-flip operator can correspond to the action of toggling a switch, however, for others, it may not, for example, the alphabets do not have semantic meaning by themselves.

We are now equipped with a symbol set consisting of (i) some set of observed states and pseudo-states and (ii) a rich (universal) set of transforms to describe other observed percepts.
As a digress, it is crucial to note that transformations can be represented as states in a higher dimension via channel-state duality~\cite{jiang2013channel}, representing dynamics as statics.
Now, can we choose an optimal encoding scheme subjective to the probabilities of the various symbols being used to describe the physical phenomena (from the set of all macroscopic percepts)?
The problem is that we do not know beforehand the ways in which the blocks will be used, i.e. the distribution of the ensemble.
Imagine operating a Lego factory and deciding how many of each block to manufacture for customers' needs.
Most often, due to lack of any other information, encoding of the base set is chosen as the standard binary encoding, (i.e., with the assumption that all blocks and initial percepts will be required with uniform probability), e.g. the ASCII code.
This is called the opcode encoding of the instruction set architecture in computers.
In this scenario (of universal computation), it can be useful to study things from the other end, i.e., what will be the resources required to represent a specific percept.
Resources are typically of two flavors: (i) the computational cost in terms of cycles (time) and memory (space), and (ii) the length of the description of the percept using the language.

The computational cost is studied in the field of computational complexity.
Problems (and thereby, their solutions as a sequence of instructions based on symbols), are classified into different classes~\cite{czoo} based on the scaling behavior of time and space with the size of the problem.
Some common ones are polynomial time (P), non-deterministic polynomial time (NP) and bounded-error quantum polynomial time (BQP).

The length of description quantifies the Kolmogorov complexity~\cite{kolmogorov1965three} or algorithmic entropy of the percept.
It is defined as $K_U(X) = \min_p\{\ell(p):U(p)=x\}$, where $\ell$ denotes the length of the (prefix-free) program $p$ on the encoding used by the universal Turing machine $U$ that outputs $x$.
Though it depends on the choice of the building blocks and their encodings, the dependence is only of an additive constant term (called the invariance theorem) which is the length of a cross-compiler to another language/automata.
Thus, it is useful to use Kolmogorov complexity to quantify the individual complexity of a string, irrespective of an ensemble.
However, finding the exact value is uncomputable.
There are many ways to approach it from the upper side (lower semi-computable), for example, via compression algorithms, minimum description length and the block decomposition method.

% \vspace{0.5em}
So far we reviewed three different notions of complexity of states:
\begin{enumerate}[nolistsep,noitemsep]
    \item Statistical complexity: Shannon entropy on an ensemble of states (given its probability distribution)
    \item Computational complexity: Space-time scaling behavior of a program to generate the state (given a language)
    \item Algorithmic complexity: Length of the program to generate the state (given a language)
\end{enumerate}
In this research, we are instead interested in the circuit complexity of a state.
Circuit complexity is related to algorithmic complexity~\cite{allender2008circuit}, which in turn is related to statistical~\cite{grunwald2004shannon} and computational complexities~\cite{fortnow2004kolmogorov}.
Computational complexities typically deal with asymptotic scaling behavior and provides lower bounds.
Though families of circuits have specific complexity class hierarchy (e.g., $AC^i$, $TC^i$, $NC^i$) it is not of much interest for this research.
We will focus on circuits with bounded size (in both space and time).
Similarly, the expected Kolmogorov complexity has been shown to correspond to the Shannon entropy~\cite{grunwald2004shannon}, though this relation is not of immediate importance to this work.
\cite{allender2008circuit} 
Kolmogorov complexity can be shown being very similar to circuit complexity under certain considerations~\cite{allender2008circuit}.
Another similar relation is that truth tables of functions with small circuit complexity has small Kolmogorov complexity.
Counting arguments relating circuit, algorithmic and statistical complexities has been suggested in \cite{toffoli2018action,lloyd2016universal} in terms of Lagrangian action.
Our research in another step in this rather niche field of understanding observed states via different perspectives.

It is important to note that most research on algorithmic information theory has been in the context of universal automata, e.g. Turing machines, lambda calculus, cellular automata, etc.
The size of the description depends on how expressive the symbols are for the transformations.
What we described so far, i.e., transformations as a relation between two states, is typically the case in the language of circuits.
Program written in more abstract logical framework allow more powerful primitives, like universal and existential quantifiers in first-order or higher-order logic.
Typically, an universal computation model demands a recursively enumerable language.
In the Chomsky hierarchy, Turing machines are more powerful than linear-bounded automata, which are inturn more powerful than push-down automata and in turn, finite-state machines~(FSM).
See \cite{sarkar2020quantum} for a comparison of these for both classical and quantum computing models.
However, for less powerful automata and language models, it is possible to derive corresponding notions~\cite{zenil2019coding} of algorithmic complexity.
This is important as programs written in Turing-complete languages eventually gets translated via the layers of the computing stack and gets executed by logic circuits.
These logic circuits are however a combination of sequential (allowing memory cells) and combinatorial logic, and can be used to simulate an FSM.
Purely combinatorial logic (not to be confused with combinatory logic, which is universal) is of even lower power than FSM.
The former is loopless and stateless, and thereby is a direct representation of the output state based on the input.
It is important to note that, program execution is typically clocked in both classical and quantum processors to prevent race-conditions, even if the circuits are purely composed of combinatorial logic elements.
Thus, resources of time and space can be defined in this setting even without tracking and accessing intermediate states.
% one can write an FSM to check if the string is all-ones or not (of arbitrary length), but that corresponds to a family of circuits, one for each length; there's no state memory of "so far all ones"
By borrowing notions from algorithmic information theory (as defined on functional programs), in this work, we study the effect of circuit complexity of Boolean/quantum combinatorial logic on state complexity.

% \textcolor{red}{Questions remaining:}
% \begin{itemize}[nolistsep,noitemsep]
%     \item Binary combinatory logic using SK-Combinators is Turing-complete, but Boolean logic gate is not, even though both is quantizer free. Why?
% \end{itemize}

\section{Landscape of circuits} \label{s:circuits}

With this background of the measures of complexity, let us now first explore the landscape of Boolean circuits.
The quantum circuit model is inspired by and is a generalization of the Boolean circuit model, so, it would be natural to start with a classical model and generalize it to the corresponding quantum formulation.

\subsection{Circuit probability of states}

Algorithmic information is typically studied for classical functions (e.g. for $\lambda$-calculus) than for combinatorial Boolean logic circuits.
We intend to study the latter.
Let us consider the space of n-bit strings.
Given a set of gates that form a Boolean circuit, we find that, all outputs are not equally likely.
This is because, while each \{circuit, input\} pair has only one output, there are many ways of generating the same outputs from multiple circuits.
In fact, we can make our circuits arbitrarily big by dummy operations like identity or two consecutive NOT-gates.

Since there are many programs, to compare two strings, instead of finding the shortest circuit to output the string, we are interested in the probability of each circuit being generated.
This is similar to the notion of the algorithmic probability~\cite{solomonoff1964formal} of the string and is defined as $M(X)=\sum_{p:U(p)=x*} 2^{-\ell(p)}$  when the prefix-free programs $p$ on the universal automata $U$ are encoded in binary.
The largest contribution to this term comes from the shortest program (i.e. the Kolmogorov complexity).
This connection between complexity and probability can be expressed as: a string which has a short program has many alternate ways of generating it and is thus more probable to get generated by a universal automaton programmed randomly.
Note that, assigning an uniform random distribution of programs for generating the algorithm probability, or the universal distribution over the entire set of strings, is not fully justified.
In \S~4.5 of \cite{sarkar2022applications} one of the authors proposed a more physically motivated `nested algorithmic probabilities' that converges to constructors.
In this work, we will start with a uniform distribution but will later generalize the implementation to allow any prior distribution.
To distinguish the usual notion of algorithmic probability of a string on an universal automata $M_U(X)$ from our case of the probability of an output string based on the distribution of equivalent circuits with varied space-time complexities, we denote our formulation of algorithmic probability as $M_{circ}(X)$.

In the original setting, $M_U(X)$ is uncomputable, as it requires running each possible program, of which there exists programs that does not halt.
However, it is lower semi-computable, and can be approximated given bounds on run-time.
One proposal to approximate is given in \cite{soler2014calculating} by running every Turing machine in a particular enumeration, and directly using the output distribution of halting Turing machines upto the bounded run-time. 
In the case of Boolean/quantum circuits, the run-time bounds are predetermined, and there is no halting problems.
Thus, $M_{circ}(s)$ for a state $s$ can be approximated by the ratio of cardinality of the sets that generate the target state from the initial state $s_0$ with the total number of circuits, as:
\begin{equation}\label{eq:mcirc}
M_{circ}(s) \approx \dfrac{|C \in \mathtt{\{gate set, max space, max time\}}, s \leftarrow C(s_0)|}{|C \in \mathtt{\{gate set, max space, max time\}}|}
\end{equation}

This can be used to estimate the quantum circuit complexity using the coding theorem by extending the relation~\cite{delahaye2012numerical} between probability and complexity to circuits as, $K_{circ}(s) = -\log M_{circ}(s)$.

\subsection{Boolean gate sets}

In the Boolean circuit form of algorithmic probability, we will consider strings of n-bits, and the probabilities of each bit string getting generated from all possible Boolean circuits on all possible inputs.
The main restriction (i.e., the output not being also uniformly random) comes from the fact that we do not have primitives (1-time step gates) for all possible Boolean functions. 
We typically use a universal gate set that can compile any other Boolean functions down to a larger number of gates from that set.
Thus in our operational implementation, we need to choose a gate set for the empirical enumeration of the circuits.

% These choices are presented in the implementation section.

Given $v$ input variables with a symbol set of size $s$, there are $s^v$ possible combinations of these inputs.
If there is a single output variable from the symbol set of size $d$, the total number of possible functions~\cite{wernick1941complete} is $d^{s^v}$.
\begin{itemize}[nolistsep,noitemsep]
    \item For 1-input Boolean algebra, i.e. when $v = 1$, $s = 2$, $d = 2$, the total number of functions are $f = 2^{2^1} = 4$.
    These functions are the $\{0,1,A,\overline{A}\}$.
    \item For 2-input Boolean algebra, i.e. when $v = 2$, $s = 2$, $d = 2$, the total number of functions are $f = 2^{2^2} = 16$.
    These are denoted by $\{0,1,A,B,\overline{A},\overline{B},A \bullet B,\overline{A \bullet B},A+B,\overline{A+B},A+\overline{B},\overline{A}+B,A \bullet\overline{B},\overline{A} \bullet B,A\oplus B,\overline{A\oplus B}\}$
\end{itemize}
A functionally complete set of logical connectives or Boolean operators can be used to express all possible truth tables by combining members of the set into a Boolean expression.
These sets can also express any Boolean SAT, or SAT-3 formula.
Some examples of such universal~\cite{sheffer1913set} sets are \{\verb|NAND|\}, \{\verb|NOR|\}, \{\verb|NOT|, \verb|AND|\}, \{\verb|NOT|, \verb|OR|\}.
These gate sets are related to each other, using the following equivalences:
\begin{itemize}[nolistsep,noitemsep]
    \item \verb|NOT(A) = NAND(A,A) = NOR(A,A)|
    \item \verb|OR(A,B) = NAND(NAND(A,A),NAND(B,B)) = NOR(NOR(A,B),NOR(A,B)) = NOT(AND(NOT(A),NOT(B)))|
    \item \verb|AND(A,B) = NAND(NAND(A,B),NAND(A,B)) = NOR(NOR(A,A),NOR(B,B)) = NOT(OR(NOT(A),NOT(B)))|
\end{itemize}
% Some other common equivalences are:
%     IMPL(A,B) = OR(NOT(A),B)
%     BICON(A,B) = AND(IMPL(A,B),IMPL(B,A))

% \section{Resource-bounded quantum programs} \label{s4}

\subsection{Quantum gate sets}

The classical formulation that maps the landscape of Boolean functions can now be generalized to include quantum gates and states.
There is a 3-input single gate in quantum logic that is universal for classical computing, the \verb|CCX| gate (also called the Toffoli gate).
It can simulate the \verb|NAND| gate via \verb|CCX(A,B,1) = (A,B,NAND(A,B))|.
Classical computation is in general an irreversible process, thus the inputs cannot be recovered from the outputs.
Quantum logic is based on unitary evolution and thus is reversible.
Additionally, quantum computations allow quantum superposition and entanglement, which are not implied in reversible computation.
The \verb|CCX| gate can simulate the entire reversible computation by simulating a \verb|Fanout| gate (or Copy gate) as \verb|CCX(A,1,0) = (A,1,A)|.
Thus, both \{\verb|NAND|, \verb|Fanout|\} and \{\verb|CCX|\} form universal gate sets for reversible computation.
The \verb|CSWAP| gate (also called the Fredkin gate) is another universal gate for reversible logic.

The generalization of reversible to quantum logic needs only one extra gate, the \verb|H| gate (Hadamard).
In principle, the real gate set composed of \{\verb|CCX|, \verb|H|\} is computationally universal~\cite{shi2002both}.
However, it needs ancilla qubits to encode the real normalization factors and complex algebra to decompose~\cite{aharonov2003simple} to arbitrary quantum unitary gates for a strong sense of universality.
Also, it is important that the effect of the \verb|NOT| gate (or, the \verb|X| gate in quantum) cannot be simulated without assuming the availability of both $\ket{0}$ and $\ket{1}$ states.
Since our enumeration of quantum programs will start will the qubits initialized to the all-zero state, we need to augment the gate set to \{\verb|X|, \verb|H|, \verb|CCX|\} to reach all binary strings as output.

The principle of algorithmic probability should also hold in the quantum setting, i.e., a uniform distribution of all possible functions and all possible input states does not imply a uniform distribution of all possible output states on the Hilbert space. 
Nielsen's geometric quantum computing~(GQC) approach~\cite{nielsen2006quantum} shows that finding optimal quantum circuits is essentially equivalent to finding the shortest path between two points in a certain curved Riemannian geometry. 
However, it is not possible to empirically visualize this, as we need to consider all possible input states and all possible unitary maps.
Studying the landscape of program synthesis requires discretizing this space for the native gate set of the target quantum processor (or the quantum compiler).
Also, the number of possible functions or processes in a quantum environment (even for a single qubit) is uncountably infinite.
Thus, choosing a universal gate set gets more pronounced in the setting of quantum control.

In a way this is easy.
It has been shown that if one can apply some Hamiltonian repeatedly to a few variables at a time one can in general affect any desired unitary time evolution on an arbitrarily large number of variables.
As a result, almost any quantum logic gate with two or more inputs is computationally universal~\cite{lloyd1995almost} in a way that copies of the gate can be wired together to effect any desired logic circuit, and to perform any desired unitary transformation on a set of quantum variables.
We call this richer counterpart to its classical cousin~\cite{wolfram2002new}, the ubiquity of quantum universality~(UQU).

How many types of quantum gates in the gate set do we need to represent this richer set of quantum unitary operators, and how many of them do we need?
Well, if we are provided with a parametric family of quantum operators, only a few types of such operators are sufficient.
The quantum Shannon decomposition~(QSD)~\cite{shende2005synthesis} provides a theoretical lower bound and asymptotic optimality for an exact decomposition of quantum unitaries using the parametric family of gates \{\verb|RY|$(\theta)$, \verb|RZ|$(\theta)$, \verb|CX|\}. 
It can be recursively applied to larger quantum circuits with the CX count scaling of $O(4^n)$.

GQC, UQU and QSD, relies on an arbitrary expressive set of gates.
This is not very practical as quantum devices are manufactured and controlled to perform operations from a predefined dictionary.
There is a subtle difference of using a finite set of operators with respect to the classical case.
Instead of the classical setting of $d^{s^v}$ being represented perfectly by a sequence of gates from the universal gate set $G$, in the quantum setting, the aim is to approximate all possible unitary operations with a sequence of gates from $G$ with a bound of the approximation quality.
This can be understood by thinking of representing all real numbers using digits of a specific numeral base.
Of course there is a trade off to taming this countably infinite space with a finite number of building blocks.
Quantum Kolmogorov complexity~(QKC)~\cite{mora2007quantum} is a measure of the information required to describe a quantum state. 
For any definition of quantum Kolmogorov complexity measuring the number of classical bits required to describe a pure quantum state, there exists a pure n-qubit state which requires exponentially many bits of description.

Nevertheless, the Solovay-Kitaev theorem~(SKT)~\cite{dawson2005solovay} allows an efficient classical algorithm for compiling an arbitrary single-qubit gate into a sequence of gates from a fixed and finite set. 
The algorithm, using a universal gate set~\cite{barenco1995elementary} (e.g., \{\verb|H|, \verb|T|, \verb|CX|\}), runs in $O(log(1/\epsilon))$ time, and produces as output a sequence of $O(log(1/\epsilon))$ quantum gates which approximates the desired quantum gate to an accuracy within $\epsilon>0$.
It can be generalized to apply to multi-qubit gates and to gates from $SU(d)$.

In retrospect, there is no foundational reason known why GQC, UQU, QSD, QKC and SKT plays out in Nature in this manner.
Yet, eventually, it allows us to sufficiently parse and explore the vast Hilbert space using an arbitrary choice of a small set of building blocks.
In the next section, we will present a formal formulation of our emuneration procedure, the results and their analysis.

\section{Implementation} \label{s:implementation}

We first describe the implementation of the classical case.
The problem is formulated as follows: given (i) $n$ bits, $b_i \in \{b_0,b_1,\dots,b_{n-1}\} = B$, (ii) an initial state for each bit $s_0(b_i)$ (typically set to $0$), (iii) a set of gates $g \in G$ (not necessarily universal), and, (iv) number of lines of QASM code $L$; find the distribution of final states given each gate is applied with probability $\frac{1}{|G|}$ at each $l \in L$.

In the quantum case, the gate set is now defined as a set of unitary gates, while the initial state over a set of $n$ qubits $Q$ is defined as $s_0(Q) := \sum_{j \in \{0,2^n-1\}} \alpha_j \ket{j}$, such that $\alpha_j \in \mathbb{C}$ and $\ket{j}$ are eigenstates of the $n$-dimensional Hilbert space in the Z-basis.

\subsection{Gate sets}

We consider the following gate sets:
\begin{enumerate}[nolistsep,noitemsep]
    \item \{\verb|CCX|\} - This set is universal for classical and reversible logic, provided both the initial states of $\ket{0}$ and $\ket{1}$ is provided. It is not practical to provide all initial states without knowing how to create one from the other. Since all gate-based quantum algorithms start from the all-$\ket{0}$ state and prepare the required initial state via gates, we will not consider this set for our enumeration. 
    % The distribution of this set is embedded within the distribution of the following gate sets, by filtering various initial state conditions.
    \item \{\verb|X|, \verb|CCX|\} - This set is universal for classical and reversible logic by starting from the all-$\ket{0}$ state.
    \item \{\verb|X|, \verb|H|, \verb|CCX|\} - This set is weakly universal under encoding and ancilla assumptions for quantum logic. The encoding, while universal, might not preserve the compututation resource complexity benefits of quantum (i.e., in the same way classical computation can also encode all quantum computation using \{\verb|NAND|, \verb|Fanout|\}). Thus, we do not consider this set for our enumeration of the quantum case. 
    \item \{\verb|H|, \verb|S|, \verb|CX|\} - The Clifford group is useful for quantum error correction. However, it is non-universal and can be efficiently simulated on classical logic~\cite{gottesman1998heisenberg}. The space of transforms on this set encoded error-correction codes and is thus useful to map. 
    \item \{\verb|H|, \verb|T|\} - This set is universal for single qubit quantum logic. However, we will consider the generalization to multi-qubit using an additional two-qubit gate in the set in the following case.
    \item \{\verb|H|, \verb|T|, \verb|CX|\} - This is universal for quantum logic.
    \item \{\verb|P(pi/4)|, \verb|RX(pi/2)|, \verb|CX|\} - The IBM native gate set is used to construct this gate set. The following relations establish the relation with the previous universal gate set: $\verb|T| = \verb|P(pi/4)|$, $\verb|X| = \verb|RX(pi/2)|$, and, $\verb|H| = e^{i\pi/2} \verb|X| \verb|Rz(pi/2)| \verb|X| = e^{i\pi/2} \verb|XTTTTX| $. We will consider additional constraints like device connectivity to apply this technique to real quantum processors. 
\end{enumerate}
Thus, in our experiments, we map the algorithmic probability of the final states for the following gate sets: (i) \{\verb|X|, \verb|CCX|\}, (ii) \{\verb|H|, \verb|S|, \verb|CX|\}, (iii) \{\verb|H|, \verb|T|, \verb|CX|\}, and (iv) \{\verb|P(pi/4)|, \verb|RX(pi/2)|, \verb|CX|\}.

\subsection{Metrics for evaluation}
We are interested in evaluating these metrics for each of the gate sets:
\begin{itemize}[nolistsep,noitemsep]
    \item Expressivity: refers to the extent to which the Hilbert space can be encoded by using an unbounded number of gates. It is not weighted by the probability as it is a characteristic of the encoding power of the gate set. We assign a $1$ to a final state if it can be expressed as starting from the initial state and applying a sequence of gates from the gate set.
    \item Reachability: refers to a bounded form of expressibility. The length of the sequence of gates must be equal to or shorter than the specified bound. This corresponds to a physical implementation rather than the power of the gate set, and characterizes the computational complexity and thereby the decoherence time of the processor. 
\end{itemize}
The expressibility is mapped primarily to understand if the reachability bound is under/over-specified.
As the value of the circuit length $L$ is gradually increased, any universal gate set will populate the full landscape of states in the expressibility criteria, and thereby remain without variation.
It is at this limit, i.e. at the first instance of full expressibility, the reachability is best understood.
The other instance we are interested in is the infinite limit of $L$, and its effect on the reachability distribution.

These experimental procedures and the comparative study of the results are presented in the following sections.

\subsection{Enumeration procedure}

We construct the experiment by constructing all possible QASM programs.
For each gate, $g_i \in G$ in the gate set, the target number of qubits $q(g_i)$ are known.
Thereafter, given $n$ qubits, all possible permutations $\mathcal{P}$ of applying the gate are enumerated in a list, i.e. $\mathcal{P}_{q(g_i)}^n$.
The total possible options for each line of QASM is $\sum_{G}  \mathcal{P}_{q(g_i)}^n $, and thus, the total possible QASM programs for $L$ lines of code length are:
\begin{equation}\label{eq:qcircs}
    \bigg[ \sum_{G}  \mathcal{P}_{q(g_i)}^n \bigg]^L
\end{equation}

Our implemented is available at \href{https://github.com/Advanced-Research-Centre/QCircScape}{github.com/Advanced-Research-Centre/QCircScape}

As an example, consider the gate set $G = \{\verb|X|, \verb|CCX|\}$, for $n = 4$ and $L = 3$.
$q(\verb|X|)=1$ and $q(\verb|CCX|)=3$.
Thus, $\mathcal{P}_{q(\verb|X|)}^4 = 4$, and $\mathcal{P}_{q(\verb|CCX|)}^4 = 24$.
Note that even if exchanging the assignment of the two controls of the Toffoli gate has the same effect, this is a symmetry property of this gate and not in general true for 3-qubit unitaries.
Thus, the description number (program id) for these cases are treated as different computational paths.
It can be appreciated that, these two options of Toffoli gates would behave very differently in present of noise characteristic of individual qubits as well as other control constraints.
The total options for each line if QASM is $28$, and thus for length $3$, the total number of programs is $28^3 = 21952$.
This is already a large number of quantum circuits to be simulated, for a small case, and gives a preview of how large the space of programs are.

By applying all possible cases, we obtain an array of size $n$ that represents the available number of transitions from a specific state to another.
The measurement basis (here, considered to be the default Z-basis), is crucial for this research.
If we consider all possible initial states of bit-strings (Z-basis state preparations) of size $n$, we obtain a $n \times n$ matrix.
This exploration of other initial states helps us to understand the asymmetry of gates over bit values (e.g., a generalization of Toffoli gates with inverted control qubits is of 4 types: $\mathtt{\bar{C}\bar{C}X, \bar{C}CX, C\bar{C}X, CCX}$).

In the classical scenario, (e.g. for \{\verb|X|, \verb|CCX|\}), this corresponds to the statistics of the number of computational paths between these two states using arrangement of gates from the set, conforming to a specified length.
For the quantum case, the statistics corresponds to the sum of probabilities of the computational paths collapsing on measurement to the target state.
Dividing the matrix by the total number of programs, gives us the fixed-length algorithmic probability of the state on each row, conditioned on the initial state.
This normalized $n \times n$ matrix is the reachability landscape.
All non-zero values corresponds to the states that are reachable by at least one route (i.e., at least one program exist to transform to that state).
This gives us the Boolean $n \times n$ expressibility matrix.

\subsection{Results}

\begin{figure}[htb]
	\centering %LBRT
	\includegraphics[clip, trim=16cm 2.5cm 13cm 4cm,width=0.7\textwidth]{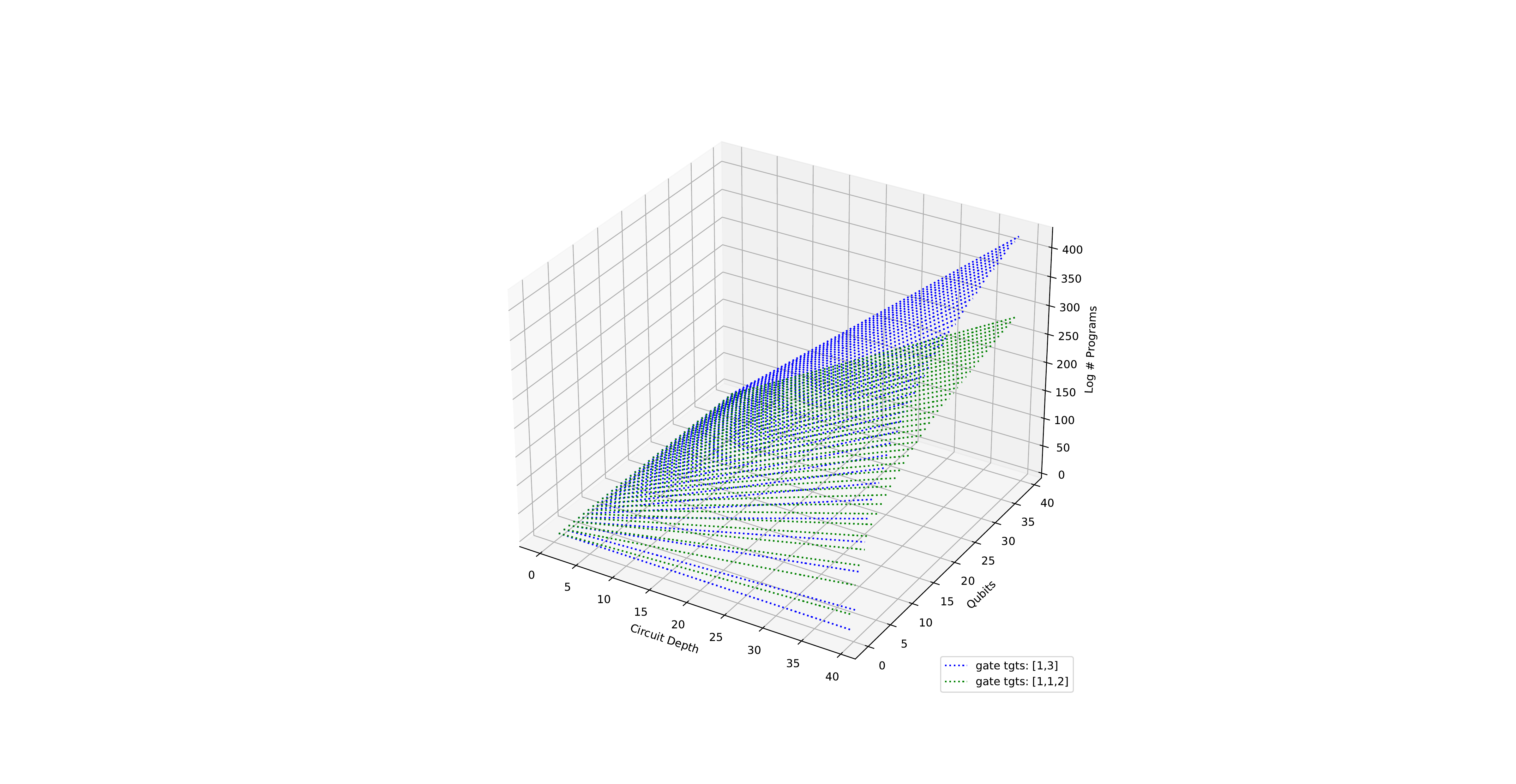}
	\caption{Growth of the number of programs with qubit count and circuit depth for two types of gate sets: (i) $[1,3]$ qubits: $\mathtt{\{X,CCX\}}$, (ii) $[1,1,2]$ qubits: $\mathtt{\{H,S,CX\}}$, $\mathtt{\{H,T,CX\}}$, $\mathtt{\{P(\pi/4), RX(\pi/2), CX\}}$ }
	\label{fig:growth}
\end{figure}

To start our enumeration, we first plot the growth of number of programs (i.e., Equation~\ref{eq:qcircs}) with qubit count and circuit depth for various gate sets.
We note that, the trend is independent of the description of the gates in the gate set.
The only information that matters is how many target qubits each gate in the set acts on.
This gives us two classes among our chosen gate sets, (i) with one $1$-qubit and one $3$-qubit gate, (ii) with two $1$-qubit and one $2$-qubit gate.
The result is plotted in Figure~\ref{fig:growth}.
We find that the permutations due to a $3$-qubit gate grows much faster that the other class.
% \textcolor{red}{ADD growth rate}

\newpage
% \subsubsection{Gate set $\mathtt{\{X,~CCX\}}$}

\begin{figure}[htb]
	\centering %LBRT
	\includegraphics[clip, trim=5cm 2.5cm 5cm 2.5cm,width=0.97\textwidth]{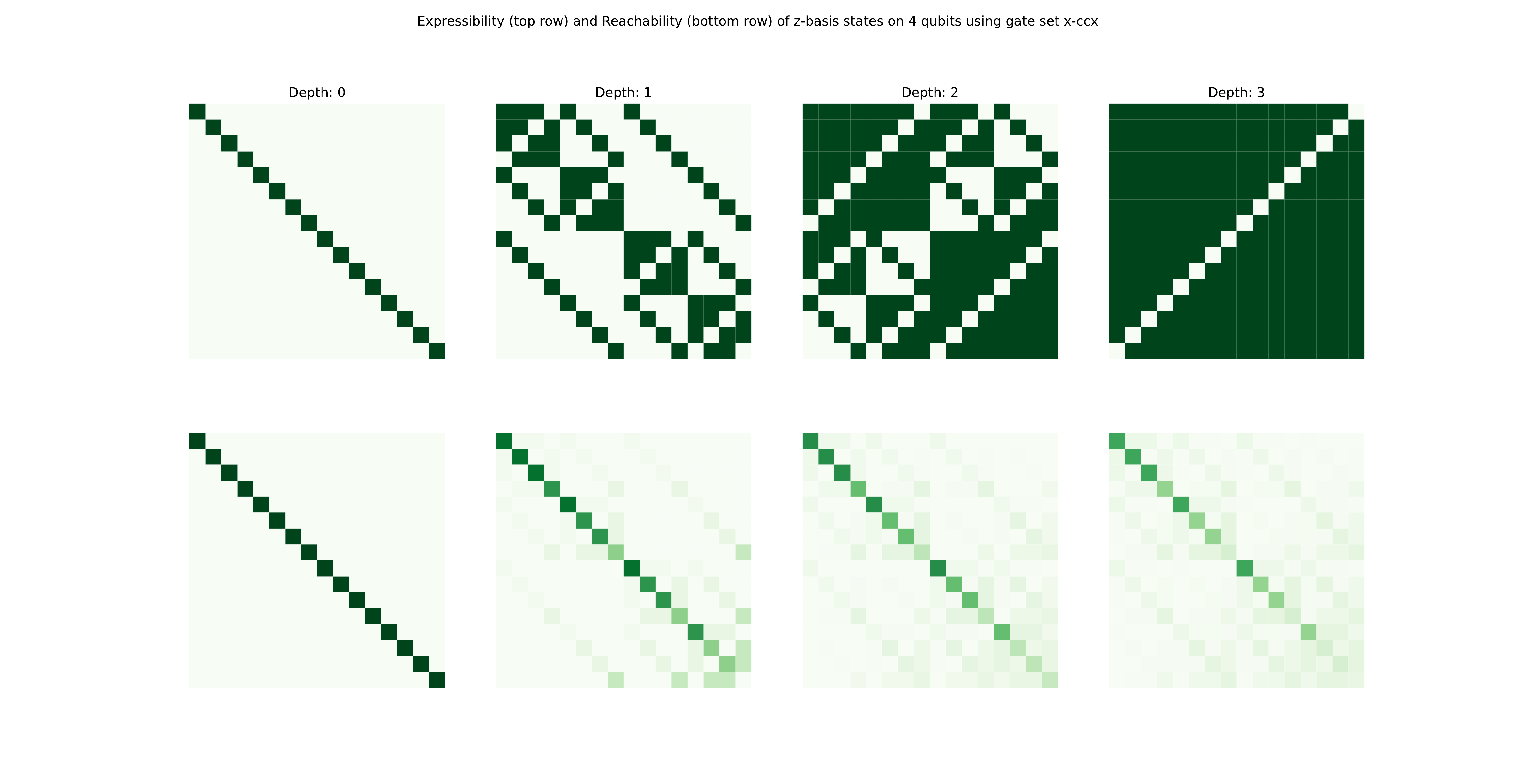}
	\caption{Expressibility and Reachability for gate set $\mathtt{\{X,CCX\}}$ on $4$ qubits and of circuit depth from 0 to 3.}
	\label{fig:gs1}
\end{figure}

\begin{figure}[htb]
	\centering %LBRT
	\includegraphics[clip, trim=5cm 2.5cm 5cm 2.5cm,width=0.97\textwidth]{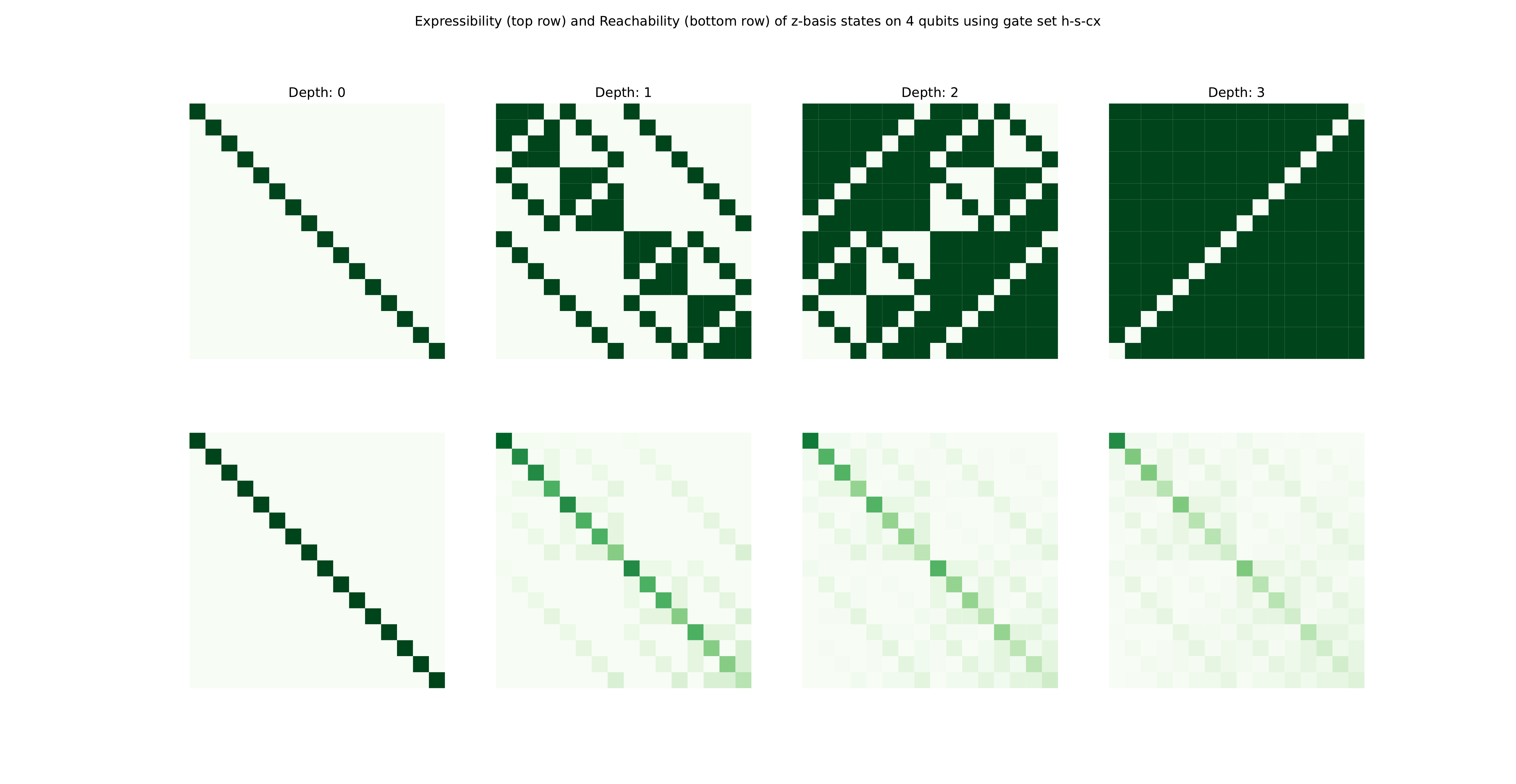}
	\caption{Expressibility and Reachability for gate set $\mathtt{\{H,S,CX\}}$ on $4$ qubits and of circuit depth from 0 to 3.}
	\label{fig:gs2}
\end{figure}

\begin{figure}[htb]
	\centering %LBRT
	\includegraphics[clip, trim=5cm 2.5cm 5cm 2.5cm,width=0.97\textwidth]{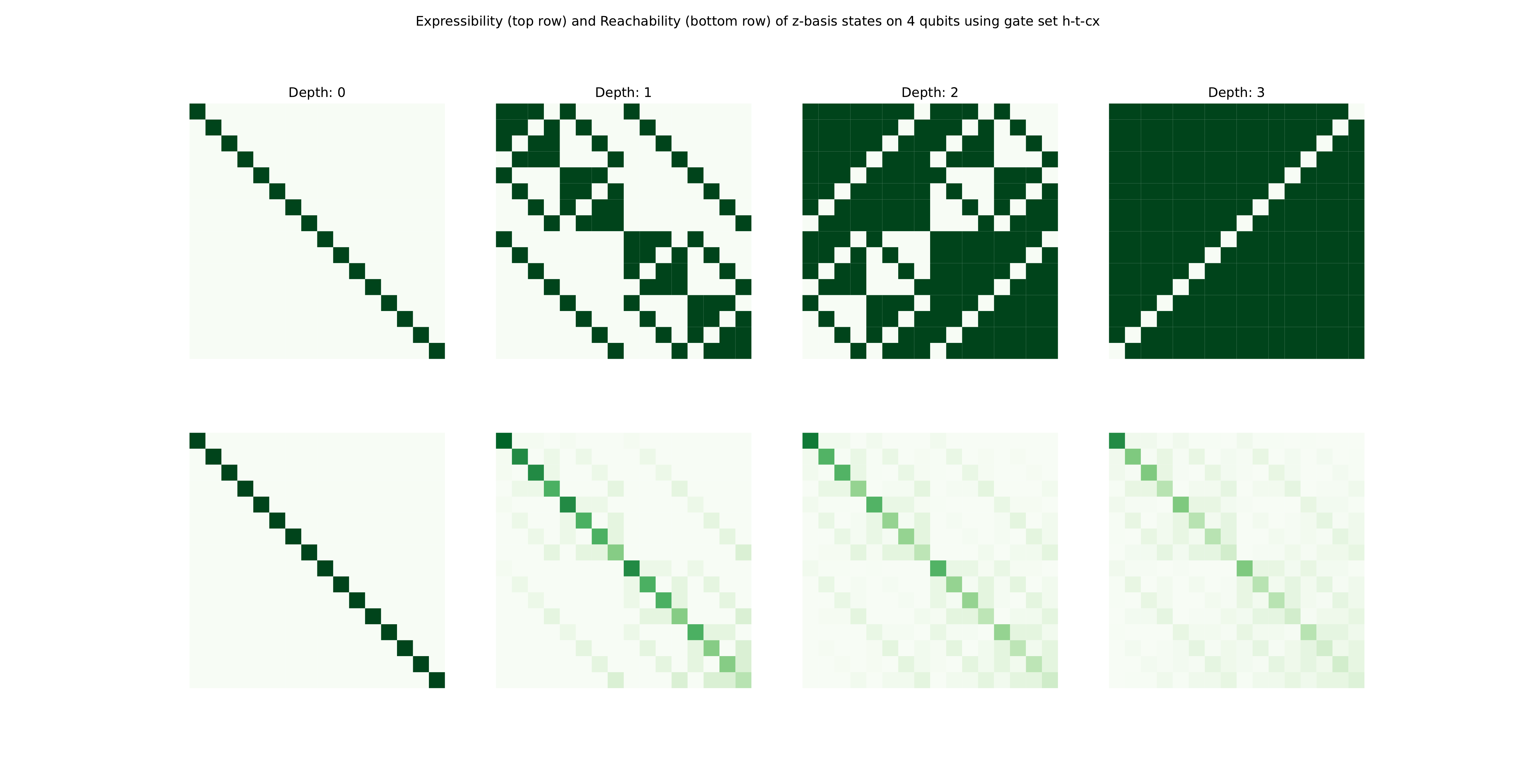}
	\caption{Expressibility and Reachability for gate set $\mathtt{\{H,T,CX\}}$ on $4$ qubits and of circuit depth from 0 to 3.}
	\label{fig:gs3}
\end{figure}

\begin{figure}[htb]
	\centering %LBRT
	\includegraphics[clip, trim=5cm 2.5cm 5cm 2.5cm,width=0.97\textwidth]{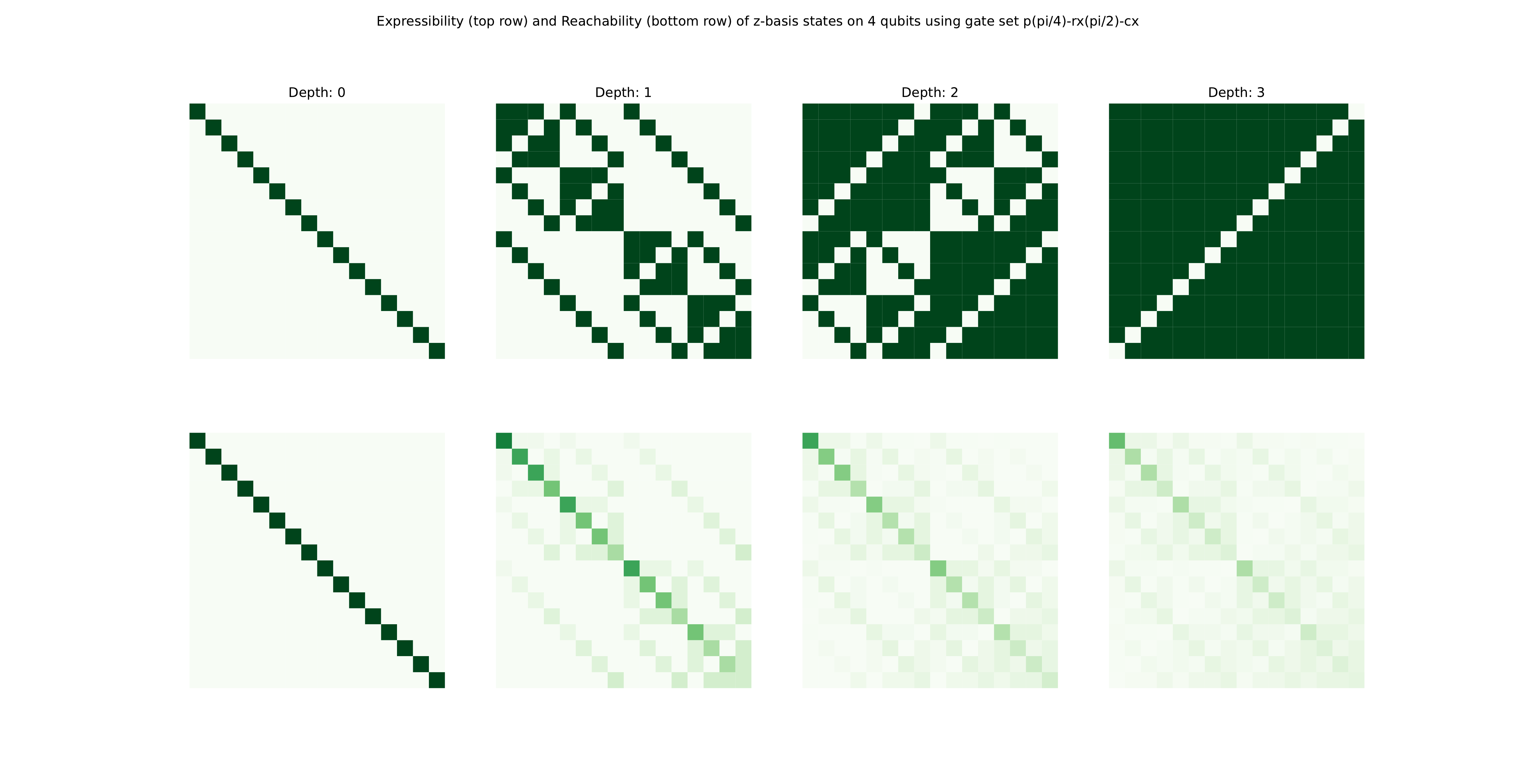}
	\caption{Expressibility and Reachability for gate set $\mathtt{\{P(\pi/4),RX(\pi/2),CX\}}$ on $4$ qubits and of circuit depth from 0 to 3.}
	\label{fig:gs4}
\end{figure}

The following figures visualizes the expressibility (top row) and reachability (bottom row) for gate set on 4 qubits with increasing depth (from 0 to 4 operations).
The gate sets we consider are $\mathtt{\{X,~CCX\}}$ (Figure~\ref{fig:gs1}), $\mathtt{\{H,~S,~CX\}}$ (Figure~\ref{fig:gs2}), $\mathtt{\{H,~T,~CX\}}$ (Figure~\ref{fig:gs3}) and $\mathtt{\{P(pi/4),~RX(pi/2),~CX\}}$ (Figure~\ref{fig:gs4}).

% \subsubsection{Gate set $\mathtt{\{H,~S,~CX\}}$}

\newpage
% \subsubsection{Gate set $\mathtt{\{H,~T,~CX\}}$}

% \subsubsection{Gate set $\mathtt{\{P(pi/4),~RX(pi/2),~CX\}}$}

\newpage
\subsection{Analysis and discussion}

Let us consider $\mathtt{maxspace} = 4$ and the classical gate set, $\mathtt{\{X,~CCX\}}$.
There are $28$ distinct possibilities for each time step.
% X0, X1, X2, X3, CCX012, CCX021, CCX102, CCX120, CCX201, CCX210, CCX013, CCX031, CCX103, CCX130, CCX301, CCX310, CCX032, CCX023, CCX302, CCX320, CCX203, CCX230, CCX312, CCX321, CCX132, CCX123, CCX231, CCX213
The reachability statistics for the Z-basis states for time step $0,1,2,3$ are:

$$R_0^\mathtt{\{X,~CCX\}} = \begin{smallmatrix}
1 & 0 & 0 & 0 & 0 & 0 & 0 & 0 & 0 & 0 & 0 & 0 & 0 & 0 & 0 & 0 \\
0 & 1 & 0 & 0 & 0 & 0 & 0 & 0 & 0 & 0 & 0 & 0 & 0 & 0 & 0 & 0 \\
0 & 0 & 1 & 0 & 0 & 0 & 0 & 0 & 0 & 0 & 0 & 0 & 0 & 0 & 0 & 0 \\
0 & 0 & 0 & 1 & 0 & 0 & 0 & 0 & 0 & 0 & 0 & 0 & 0 & 0 & 0 & 0 \\
0 & 0 & 0 & 0 & 1 & 0 & 0 & 0 & 0 & 0 & 0 & 0 & 0 & 0 & 0 & 0 \\
0 & 0 & 0 & 0 & 0 & 1 & 0 & 0 & 0 & 0 & 0 & 0 & 0 & 0 & 0 & 0 \\
0 & 0 & 0 & 0 & 0 & 0 & 1 & 0 & 0 & 0 & 0 & 0 & 0 & 0 & 0 & 0 \\
0 & 0 & 0 & 0 & 0 & 0 & 0 & 1 & 0 & 0 & 0 & 0 & 0 & 0 & 0 & 0 \\
0 & 0 & 0 & 0 & 0 & 0 & 0 & 0 & 1 & 0 & 0 & 0 & 0 & 0 & 0 & 0 \\
0 & 0 & 0 & 0 & 0 & 0 & 0 & 0 & 0 & 1 & 0 & 0 & 0 & 0 & 0 & 0 \\
0 & 0 & 0 & 0 & 0 & 0 & 0 & 0 & 0 & 0 & 1 & 0 & 0 & 0 & 0 & 0 \\
0 & 0 & 0 & 0 & 0 & 0 & 0 & 0 & 0 & 0 & 0 & 1 & 0 & 0 & 0 & 0 \\
0 & 0 & 0 & 0 & 0 & 0 & 0 & 0 & 0 & 0 & 0 & 0 & 1 & 0 & 0 & 0 \\
0 & 0 & 0 & 0 & 0 & 0 & 0 & 0 & 0 & 0 & 0 & 0 & 0 & 1 & 0 & 0 \\
0 & 0 & 0 & 0 & 0 & 0 & 0 & 0 & 0 & 0 & 0 & 0 & 0 & 0 & 1 & 0 \\
0 & 0 & 0 & 0 & 0 & 0 & 0 & 0 & 0 & 0 & 0 & 0 & 0 & 0 & 0 & 1
\end{smallmatrix}$$

$$R_1^\mathtt{\{X,~CCX\}} = \begin{smallmatrix}
24 &  1 &  1 &  0 &  1 &  0 &  0 &  0 &  1 &  0 &  0 &  0 &  0 &  0 &  0 &  0 \\
 1 & 24 &  0 &  1 &  0 &  1 &  0 &  0 &  0 &  1 &  0 &  0 &  0 &  0 &  0 &  0 \\
 1 &  0 & 24 &  1 &  0 &  0 &  1 &  0 &  0 &  0 &  1 &  0 &  0 &  0 &  0 &  0 \\
 0 &  1 &  1 & 20 &  0 &  0 &  0 &  3 &  0 &  0 &  0 &  3 &  0 &  0 &  0 &  0 \\
 1 &  0 &  0 &  0 & 24 &  1 &  1 &  0 &  0 &  0 &  0 &  0 &  1 &  0 &  0 &  0 \\
 0 &  1 &  0 &  0 &  1 & 20 &  0 &  3 &  0 &  0 &  0 &  0 &  0 &  3 &  0 &  0 \\
 0 &  0 &  1 &  0 &  1 &  0 & 20 &  3 &  0 &  0 &  0 &  0 &  0 &  0 &  3 &  0 \\
 0 &  0 &  0 &  3 &  0 &  3 &  3 & 12 &  0 &  0 &  0 &  0 &  0 &  0 &  0 &  7 \\
 1 &  0 &  0 &  0 &  0 &  0 &  0 &  0 & 24 &  1 &  1 &  0 &  1 &  0 &  0 &  0 \\
 0 &  1 &  0 &  0 &  0 &  0 &  0 &  0 &  1 & 20 &  0 &  3 &  0 &  3 &  0 &  0 \\
 0 &  0 &  1 &  0 &  0 &  0 &  0 &  0 &  1 &  0 & 20 &  3 &  0 &  0 &  3 &  0 \\
 0 &  0 &  0 &  3 &  0 &  0 &  0 &  0 &  0 &  3 &  3 & 12 &  0 &  0 &  0 &  7 \\
 0 &  0 &  0 &  0 &  1 &  0 &  0 &  0 &  1 &  0 &  0 &  0 & 20 &  3 &  3 &  0 \\
 0 &  0 &  0 &  0 &  0 &  3 &  0 &  0 &  0 &  3 &  0 &  0 &  3 & 12 &  0 &  7 \\
 0 &  0 &  0 &  0 &  0 &  0 &  3 &  0 &  0 &  0 &  3 &  0 &  3 &  0 & 12 &  7 \\
 0 &  0 &  0 &  0 &  0 &  0 &  0 &  7 &  0 &  0 &  0 &  7 &  0 &  7 &  7 &  0
\end{smallmatrix}$$

$$R_2^\mathtt{\{X,~CCX\}} = \begin{smallmatrix}
580 &  48 &  48 &   2 &  48 &   2 &   2 &   0 &  48 &   2 &   2 &   0 &   2 &   0 &   0 &   0 \\
 48 & 580 &   2 &  44 &   2 &  44 &   0 &   6 &   2 &  44 &   0 &   6 &   0 &   6 &   0 &   0 \\
 48 &   2 & 580 &  44 &   2 &   0 &  44 &   6 &   2 &   0 &  44 &   6 &   0 &   0 &   6 &   0 \\
  2 &  44 &  44 & 420 &   0 &  10 &  10 &  96 &   0 &  10 &  10 &  96 &   0 &   0 &   0 &  42 \\
 48 &   2 &   2 &   0 & 580 &  44 &  44 &   6 &   2 &   0 &   0 &   0 &  44 &   6 &   6 &   0 \\
  2 &  44 &   0 &  10 &  44 & 420 &  10 &  96 &   0 &  10 &   0 &   0 &  10 &  96 &   0 &  42 \\
  2 &   0 &  44 &  10 &  44 &  10 & 420 &  96 &   0 &   0 &  10 &   0 &  10 &   0 &  96 &  42 \\
  0 &   6 &   6 &  96 &   6 &  96 &  96 & 220 &   0 &   0 &   0 &  58 &   0 &  58 &  58 &  84 \\
 48 &   2 &   2 &   0 &   2 &   0 &   0 &   0 & 580 &  44 &  44 &   6 &  44 &   6 &   6 &   0 \\
  2 &  44 &   0 &  10 &   0 &  10 &   0 &   0 &  44 & 420 &  10 &  96 &  10 &  96 &   0 &  42 \\
  2 &   0 &  44 &  10 &   0 &   0 &  10 &   0 &  44 &  10 & 420 &  96 &  10 &   0 &  96 &  42 \\
  0 &   6 &   6 &  96 &   0 &   0 &   0 &  58 &   6 &  96 &  96 & 220 &   0 &  58 &  58 &  84 \\
  2 &   0 &   0 &   0 &  44 &  10 &  10 &   0 &  44 &  10 &  10 &   0 & 420 &  96 &  96 &  42 \\
  0 &   6 &   0 &   0 &   6 &  96 &   0 &  58 &   6 &  96 &   0 &  58 &  96 & 220 &  58 &  84 \\
  0 &   0 &   6 &   0 &   6 &   0 &  96 &  58 &   6 &   0 &  96 &  58 &  96 &  58 & 220 &  84 \\
  0 &   0 &   0 &  42 &   0 &  42 &  42 &  84 &   0 &  42 &  42 &  84 &  42 &  84 &  84 & 196
\end{smallmatrix}$$

$$R_3^\mathtt{\{X,~CCX\}} = \begin{smallmatrix}
14112 &  1738 &  1738 &   136 &  1738 &   136 &   136 &    18 &  1738 &   136 &   136 &    18 &   136 &    18 &    18 &     0 \\
 1738 & 14100 &   140 &  1498 &   140 &  1498 &    22 &   336 &   140 &  1498 &    22 &   336 &    22 &   336 &     0 &   126 \\
 1738 &   140 & 14100 &  1498 &   140 &    22 &  1498 &   336 &   140 &    22 &  1498 &   336 &    22 &     0 &   336 &   126 \\
  136 &  1498 &  1498 &  9064 &    22 &   532 &   532 &  2766 &    22 &   532 &   532 &  2766 &     0 &   354 &   354 &  1344 \\
 1738 &   140 &   140 &    22 & 14100 &  1498 &  1498 &   336 &   140 &    22 &    22 &     0 &  1498 &   336 &   336 &   126 \\
  136 &  1498 &    22 &   532 &  1498 &  9064 &   532 &  2766 &    22 &   532 &     0 &   354 &   532 &  2766 &   354 &  1344 \\
  136 &    22 &  1498 &   532 &  1498 &   532 &  9064 &  2766 &    22 &     0 &   532 &   354 &   532 &   354 &  2766 &  1344 \\
   18 &   336 &   336 &  2766 &   336 &  2766 &  2766 &  4092 &     0 &   354 &   354 &  1572 &   354 &  1572 &  1572 &  2758 \\
 1738 &   140 &   140 &    22 &   140 &    22 &    22 &     0 & 14100 &  1498 &  1498 &   336 &  1498 &   336 &   336 &   126 \\
  136 &  1498 &    22 &   532 &    22 &   532 &     0 &   354 &  1498 &  9064 &   532 &  2766 &   532 &  2766 &   354 &  1344 \\
  136 &    22 &  1498 &   532 &    22 &     0 &   532 &   354 &  1498 &   532 &  9064 &  2766 &   532 &   354 &  2766 &  1344 \\
   18 &   336 &   336 &  2766 &     0 &   354 &   354 &  1572 &   336 &  2766 &  2766 &  4092 &   354 &  1572 &  1572 &  2758 \\
  136 &    22 &    22 &     0 &  1498 &   532 &   532 &   354 &  1498 &   532 &   532 &   354 &  9064 &  2766 &  2766 &  1344 \\
   18 &   336 &     0 &   354 &   336 &  2766 &   354 &  1572 &   336 &  2766 &   354 &  1572 &  2766 &  4092 &  1572 &  2758 \\
   18 &     0 &   336 &   354 &   336 &   354 &  2766 &  1572 &   336 &   354 &  2766 &  1572 &  2766 &  1572 &  4092 &  2758 \\
    0 &   126 &   126 &  1344 &   126 &  1344 &  1344 &  2758 &   126 &  1344 &  1344 &  2758 &  1344 &  2758 &  2758 &  2352 
\end{smallmatrix}$$

The Reachability matrix plotted in the figures is normalized by the total possible options for each line of code (i.e., the sum for any row), for example, in this gate set, $28$ for $R_1$, $784$ for $R_2$ and $21952$ for $R_3$.
The Expressibility matrix is a binary matrix representation of non-zero elements of $R$.
It is obvious that $R_0$ only has the diagonal elements, as there is only one way to create a circuit with no gates, and they effectively maps the state to itself.
However, using $R_1$, all other $R_l$ can be directly derived by matrix exponentiation by:
\begin{equation}\label{eq:hopc}
R_l^G = (R_1^G)^l \qquad G = \mathtt{\{X,CCX\}}
\end{equation}
This seems obvious from a network theory perspective, where $n$-hop neighbors can be found by exponentiation the adjacency matrix.
Techniques from graph theory can be used to study quantum circuit~\cite{bandic2022interaction}, but that is fascinating is that it can predict the distribution of states without executing each circuit.
In our enumeration, we considered equal weights for each gates.
Machine learning models can be trained on quantum programs to find the distribution of quantum gates on realistic quantum algorithms, and thereby can be used to study the kind of states most quantum algorithms generate.

Let us now develop a notion of $M_{circ}$.
In the algorithmic probability formulation of prefix-free programs, the convergence to a semi-measure is based on the notion that the infinite sum, $\lim_{i \rightarrow \inf} \sum_{i} 2^{-i} = \dfrac{1}{2} + \dfrac{1}{4} + \dfrac{1}{8} + \dots = 1$.
In our case, we need to maintain that the contributions from subsequent lines of code decrease in a similar manner.
This gives us the following Equation~\ref{eq:mcircgn} for $M_{circ}$ given a specific gate set and system size.

\begin{equation}\label{eq:mcircgn}
M_{circ}^{G,n} = \lim_{L \rightarrow \inf} \sum_{i=1}^{L}  \bigg\{ 2^{-i}* R_i^G * \big[ \sum_{g_i \in G}  \mathcal{P}_{q(g_i)}^n \big]^{-i} \bigg\}
\end{equation}

Next, we analyze the quantum universal gate set $\mathtt{\{H,~T,~CX\}}$.
We need to be careful that now \begin{equation}\label{eq:hopq}
R_l^G \ne (R_1^G)^l \qquad G = \mathtt{\{H, T, CCX\}}
\end{equation}
We find the following values:

$$R_1^\mathtt{\{H,~T,~CCX\}} = \begin{smallmatrix}
18.  &  0.5 &  0.5 &  0.  &  0.5 &  0.  &  0.  & 0.  & 0.5 &  0.  &  0.  & 0.  &  0.  & 0.  & 0.  & 0.  \\
 0.5 & 15.  &  0.  &  1.5 &  0.  &  1.5 &  0.  & 0.  & 0.  &  1.5 &  0.  & 0.  &  0.  & 0.  & 0.  & 0.  \\
 0.5 &  0.  & 15.  &  1.5 &  0.  &  0.  &  1.5 & 0.  & 0.  &  0.  &  1.5 & 0.  &  0.  & 0.  & 0.  & 0.  \\
 0.  &  1.5 &  1.5 & 12.  &  0.  &  0.  &  0.  & 2.5 & 0.  &  0.  &  0.  & 2.5 &  0.  & 0.  & 0.  & 0.  \\
 0.5 &  0.  &  0.  &  0.  & 15.  &  1.5 &  1.5 & 0.  & 0.  &  0.  &  0.  & 0.  &  1.5 & 0.  & 0.  & 0.  \\
 0.  &  1.5 &  0.  &  0.  &  1.5 & 12.  &  0.  & 2.5 & 0.  &  0.  &  0.  & 0.  &  0.  & 2.5 & 0.  & 0.  \\
 0.  &  0.  &  1.5 &  0.  &  1.5 &  0.  & 12.  & 2.5 & 0.  &  0.  &  0.  & 0.  &  0.  & 0.  & 2.5 & 0.  \\
 0.  &  0.  &  0.  &  2.5 &  0.  &  2.5 &  2.5 & 9.  & 0.  &  0.  &  0.  & 0.  &  0.  & 0.  & 0.  & 3.5 \\
 0.5 &  0.  &  0.  &  0.  &  0.  &  0.  &  0.  & 0.  &15.  &  1.5 &  1.5 & 0.  &  1.5 & 0.  & 0.  & 0.  \\
 0.  &  1.5 &  0.  &  0.  &  0.  &  0.  &  0.  & 0.  & 1.5 & 12.  &  0.  & 2.5 &  0.  & 2.5 & 0.  & 0.  \\
 0.  &  0.  &  1.5 &  0.  &  0.  &  0.  &  0.  & 0.  & 1.5 &  0.  & 12.  & 2.5 &  0.  & 0.  & 2.5 & 0.  \\
 0.  &  0.  &  0.  &  2.5 &  0.  &  0.  &  0.  & 0.  & 0.  &  2.5 &  2.5 & 9.  &  0.  & 0.  & 0.  & 3.5 \\
 0.  &  0.  &  0.  &  0.  &  1.5 &  0.  &  0.  & 0.  & 1.5 &  0.  &  0.  & 0.  & 12.  & 2.5 & 2.5 & 0.  \\
 0.  &  0.  &  0.  &  0.  &  0.  &  2.5 &  0.  & 0.  & 0.  &  2.5 &  0.  & 0.  &  2.5 & 9.  & 0.  & 3.5 \\
 0.  &  0.  &  0.  &  0.  &  0.  &  0.  &  2.5 & 0.  & 0.  &  0.  &  2.5 & 0.  &  2.5 & 0.  & 9.  & 3.5 \\
 0.  &  0.  &  0.  &  0.  &  0.  &  0.  &  0.  & 3.5 & 0.  &  0.  &  0.  & 3.5 &  0.  & 3.5 & 3.5 & 6. 
\end{smallmatrix}$$   

$$R_2^\mathtt{\{H,~T,~CCX\}} = \begin{smallmatrix}
327.  &  16.  &  16.  &   1.5 &  16.  &   1.5 &   1.5 &   0.  &  16.  &   1.5 &   1.5 &   0.  &   1.5 &   0.  &   0.  &  0.  \\
 16.  & 234.  &   2.5 &  40.  &   2.5 &  40.  &   0.  &   7.5 &   2.5 &  40.  &   0.  &   7.5 &   0.  &   7.5 &   0.  &  0.  \\
 16.  &   2.5 & 234.  &  40.  &   2.5 &   0.  &  40.  &   7.5 &   2.5 &   0.  &  40.  &   7.5 &   0.  &   0.  &   7.5 &  0.  \\
  1.5 &  40.  &  40.  & 163.  &   0.  &   8.5 &   8.5 &  52.  &   0.  &   8.5 &   8.5 &  52.  &   0.  &   0.  &   0.  & 17.5 \\
 16.  &   2.5 &   2.5 &   0.  & 234.  &  40.  &  40.  &   7.5 &   2.5 &   0.  &   0.  &   0.  &  40.  &   7.5 &   7.5 &  0.  \\
  1.5 &  40.  &   0.  &   8.5 &  40.  & 163.  &   8.5 &  52.  &   0.  &   8.5 &   0.  &   0.  &   8.5 &  52.  &   0.  & 17.5 \\
  1.5 &   0.  &  40.  &   8.5 &  40.  &   8.5 & 163.  &  52.  &   0.  &   0.  &   8.5 &   0.  &   8.5 &   0.  &  52.  & 17.5 \\
  0.  &   7.5 &   7.5 &  52.  &   7.5 &  52.  &  52.  & 114.  &   0.  &   0.  &   0.  &  18.5 &   0.  &  18.5 &  18.5 & 52.  \\
 16.  &   2.5 &   2.5 &   0.  &   2.5 &   0.  &   0.  &   0.  & 234.  &  40.  &  40.  &   7.5 &  40.  &   7.5 &   7.5 &  0.  \\
  1.5 &  40.  &   0.  &   8.5 &   0.  &   8.5 &   0.  &   0.  &  40.  & 163.  &   8.5 &  52.  &   8.5 &  52.  &   0.  & 17.5 \\
  1.5 &   0.  &  40.  &   8.5 &   0.  &   0.  &   8.5 &   0.  &  40.  &   8.5 & 163.  &  52.  &   8.5 &   0.  &  52.  & 17.5 \\
  0.  &   7.5 &   7.5 &  52.  &   0.  &   0.  &   0.  &  18.5 &   7.5 &  52.  &  52.  & 114.  &   0.  &  18.5 &  18.5 & 52.  \\
  1.5 &   0.  &   0.  &   0.  &  40.  &   8.5 &   8.5 &   0.  &  40.  &   8.5 &   8.5 &   0.  & 163.  &  52.  &  52.  & 17.5 \\
  0.  &   7.5 &   0.  &   0.  &   7.5 &  52.  &   0.  &  18.5 &   7.5 &  52.  &   0.  &  18.5 &  52.  & 114.  &  18.5 & 52.  \\
  0.  &   0.  &   7.5 &   0.  &   7.5 &   0.  &  52.  &  18.5 &   7.5 &   0.  &  52.  &  18.5 &  52.  &  18.5 & 114.  & 52.  \\
  0.  &   0.  &   0.  &  17.5 &   0.  &  17.5 &  17.5 &  52.  &   0.  &  17.5 &  17.5 &  52.  &  17.5 &  52.  &  52.  & 87.
\end{smallmatrix}$$     

$$(R_1^\mathtt{\{H,~T,~CCX\}})^2= \begin{smallmatrix}
325.  &  16.5 &  16.5 &   1.5 &  16.5 &   1.5 &   1.5 &   0.  &  16.5 &   1.5 &   1.5 &   0.  &   1.5 &   0.  &   0.   &  0.  \\
 16.5 & 232.  &   2.5 &  40.5 &   2.5 &  40.5 &   0.  &   7.5 &   2.5 &  40.5 &   0.  &   7.5 &   0.  &   7.5 &   0.   &  0.  \\
 16.5 &   2.5 & 232.  &  40.5 &   2.5 &   0.  &  40.5 &   7.5 &   2.5 &   0.  &  40.5 &   7.5 &   0.  &   0.  &   7.5  &  0.  \\
  1.5 &  40.5 &  40.5 & 161.  &   0.  &   8.5 &   8.5 &  52.5 &   0.  &   8.5 &   8.5 &  52.5 &   0.  &   0.  &   0.   & 17.5 \\
 16.5 &   2.5 &   2.5 &   0.  & 232.  &  40.5 &  40.5 &   7.5 &   2.5 &   0.  &   0.  &   0.  &  40.5 &   7.5 &   7.5  &  0.  \\
  1.5 &  40.5 &   0.  &   8.5 &  40.5 & 161.  &   8.5 &  52.5 &   0.  &   8.5 &   0.  &   0.  &   8.5 &  52.5 &   0.   & 17.5 \\
  1.5 &   0.  &  40.5 &   8.5 &  40.5 &   8.5 & 161.  &  52.5 &   0.  &   0.  &   8.5 &   0.  &   8.5 &   0.  &  52.5  & 17.5 \\
  0.  &   7.5 &   7.5 &  52.5 &   7.5 &  52.5 &  52.5 & 112.  &   0.  &   0.  &   0.  &  18.5 &   0.  &  18.5 &  18.5  & 52.5 \\
 16.5 &   2.5 &   2.5 &   0.  &   2.5 &   0.  &   0.  &   0.  & 232.  &  40.5 &  40.5 &   7.5 &  40.5 &   7.5 &   7.5  &  0.  \\
  1.5 &  40.5 &   0.  &   8.5 &   0.  &   8.5 &   0.  &   0.  &  40.5 & 161.  &   8.5 &  52.5 &   8.5 &  52.5 &   0.   & 17.5 \\
  1.5 &   0.  &  40.5 &   8.5 &   0.  &   0.  &   8.5 &   0.  &  40.5 &   8.5 & 161.  &  52.5 &   8.5 &   0.  &  52.5  & 17.5 \\
  0.  &   7.5 &   7.5 &  52.5 &   0.  &   0.  &   0.  &  18.5 &   7.5 &  52.5 &  52.5 & 112.  &   0.  &  18.5 &  18.5  & 52.5 \\
  1.5 &   0.  &   0.  &   0.  &  40.5 &   8.5 &   8.5 &   0.  &  40.5 &   8.5 &   8.5 &   0.  & 161.  &  52.5 &  52.5  & 17.5 \\
  0.  &   7.5 &   0.  &   0.  &   7.5 &  52.5 &   0.  &  18.5 &   7.5 &  52.5 &   0.  &  18.5 &  52.5 & 112.  &  18.5  & 52.5 \\
  0.  &   0.  &   7.5 &   0.  &   7.5 &   0.  &  52.5 &  18.5 &   7.5 &   0.  &  52.5 &  18.5 &  52.5 &  18.5 & 112.   & 52.5 \\
  0.  &   0.  &   0.  &  17.5 &   0.  &  17.5 &  17.5 &  52.5 &   0.  &  17.5 &  17.5 &  52.5 &  17.5 &  52.5 &  52.5  & 85. 
\end{smallmatrix}$$ 

Let us understand why this is the case.
If we start with the state $\ket{0}$, and apply the Hadamard gate, we get the state $\dfrac{1}{\sqrt{2}}\ket{0}+\dfrac{1}{\sqrt{2}}\ket{1}$.
However, in terms of probability, this gets translated in the normalized reachability matrix as $0.5\ket{0}+0.5\ket{1}$.
Now, when, another Hadamard is applied to this state, the state evolves back to $\ket{0}$, while the reachability redistributed the state and remains $0.5\ket{0}+0.5\ket{1}$.
Note that similar situation would arise also in the classical gate if we allow measurements in non-computational basis.
In essence, this is due to the fact that complex amplitudes can destructively interfere while probabilities cannot - one of the core features~\cite{renou2021quantum} that is responsible for quantum speedup.

\begin{figure}[htb]
     \centering
     \begin{subfigure}[b]{0.5\textwidth}
         \centering
         \includegraphics[clip, trim=1cm 1cm 1cm 1cm,width=0.65\linewidth]{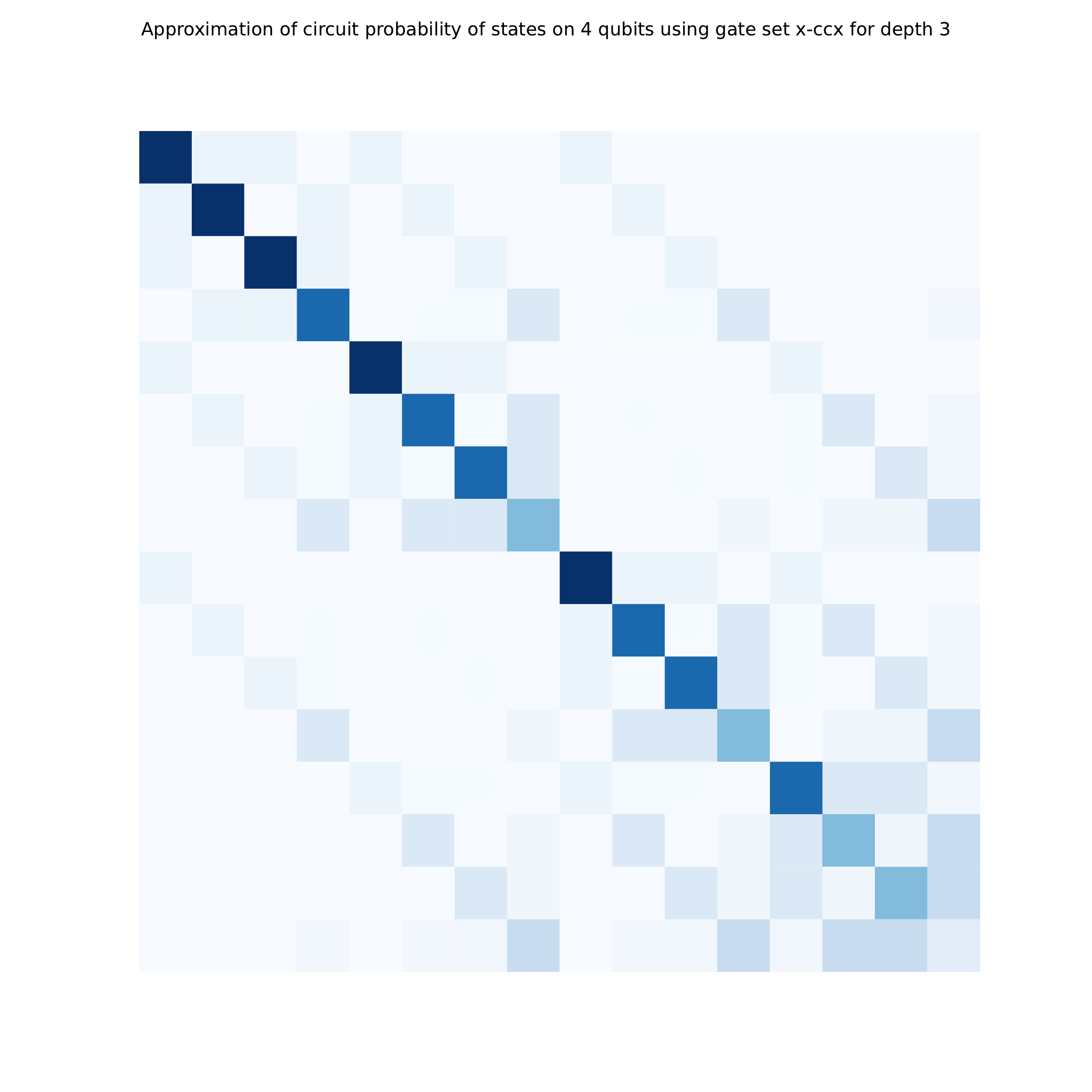}
         \caption{}
         \label{fig:mcirc1}
     \end{subfigure}%
     \begin{subfigure}[b]{0.5\textwidth}
         \centering
         \includegraphics[clip, trim=1cm 1cm 1cm 1cm,width=0.65\linewidth]{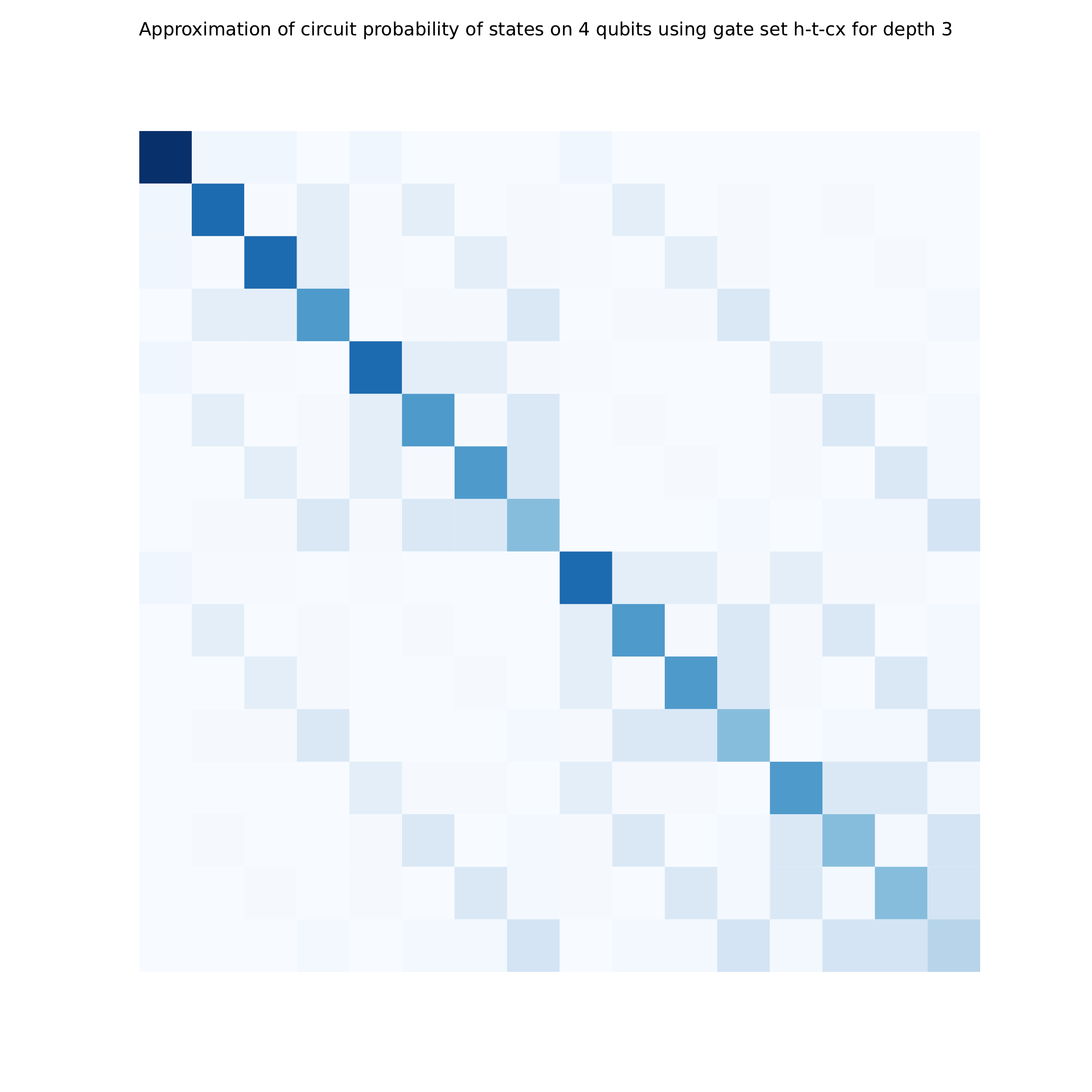}
         \caption{}
         \label{fig:mcirc2}
     \end{subfigure}
    \caption{Approximation of circuit probability of states on $4$ qubits for $L=3$ using two gate sets (a) $\mathtt{\{X,~CCX\}}$ and (b) $\mathtt{\{H,~T,~CX\}}$}
    \label{fig:mcirc}
\end{figure}

In the quantum gate set scenario, the same definition of $M_{circ}$ still hold, however it is no longer computable in linear time, but can be approximated like algorithmic probability.
We obtain the $M_{circ}$ approximated for $L=3$ for the gate sets $\mathtt{\{X,~CCX\}}$) and $\mathtt{\{H,~T,~CX\}}$ as shown in Figure~\ref{fig:mcirc}.

% \newpage 

% \begin{figure}[htb]
% 	\centering %LBRT
% 	\includegraphics[clip, trim=5cm 2.5cm 5cm 2.5cm,width=0.97\textwidth]{figures/plot_Q-5_L-3_GS-x-ccx_IBM-t.pdf}
% 	\caption{Expressibility and Reachability for gate set $\mathtt{\{P(\pi/4),RX(\pi/2),CX\}}$ on $5$ qubits and of circuit depth from 0 to 3 on the IBM T-topology with no noise.}
% 	\label{fig:gs4-t}
% \end{figure}

\newpage

\begin{figure}[thb]
	\centering %LBRT
	\includegraphics[clip, trim=5cm 2.5cm 5cm 2.5cm,width=0.96\textwidth]{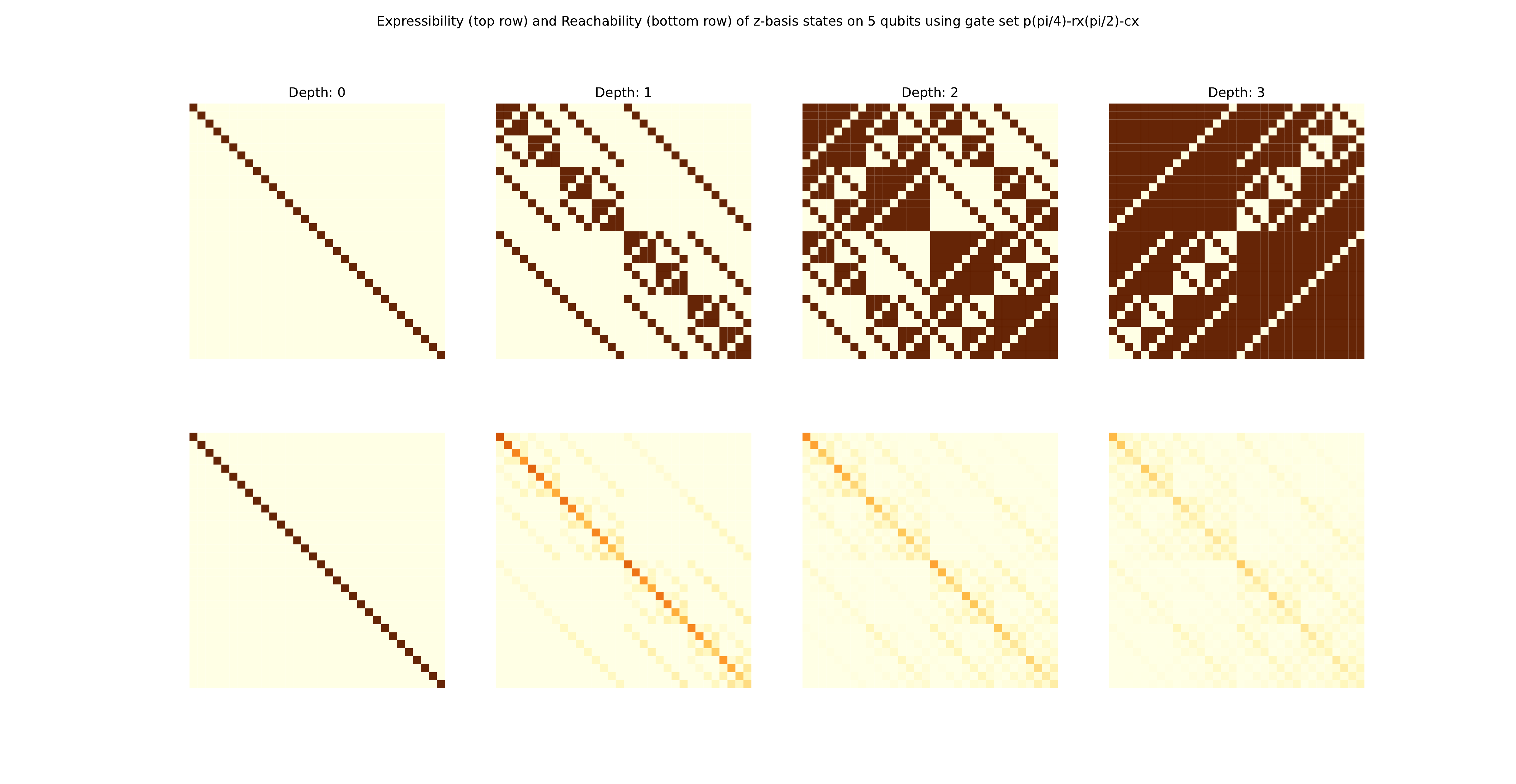}
	\caption{Expressibility and Reachability for gate set $\mathtt{\{P(\pi/4),RX(\pi/2),CX\}}$ on $5$ qubits and of circuit depth from 0 to 3 on the IBM T-topology.}
	\label{fig:gs4-t}
\end{figure}

\begin{figure}[h!]
	\centering %LBRT
	\includegraphics[clip, trim=5cm 2.5cm 5cm 2.5cm,width=0.96\textwidth]{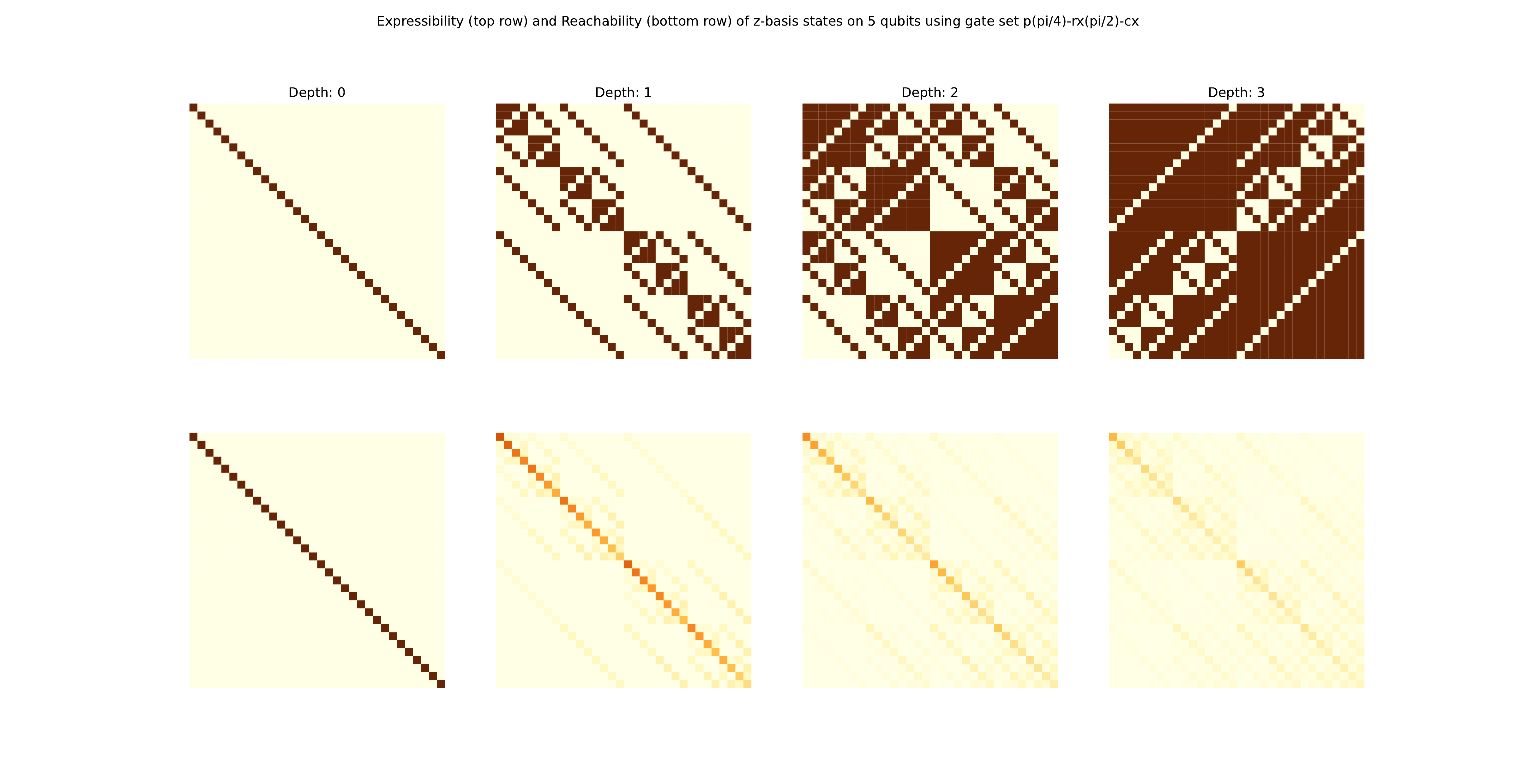}
	\caption{Expressibility and Reachability for gate set $\mathtt{\{P(\pi/4),RX(\pi/2),CX\}}$ on $5$ qubits and of circuit depth from 0 to 3 on the IBM L-topology.}
	\label{fig:gs4-l}
\end{figure}

\begin{figure}[htb]
	\centering %LBRT
	\includegraphics[clip, trim=3.5cm 1.5cm 3.5cm 1cm,width=0.95\textwidth]{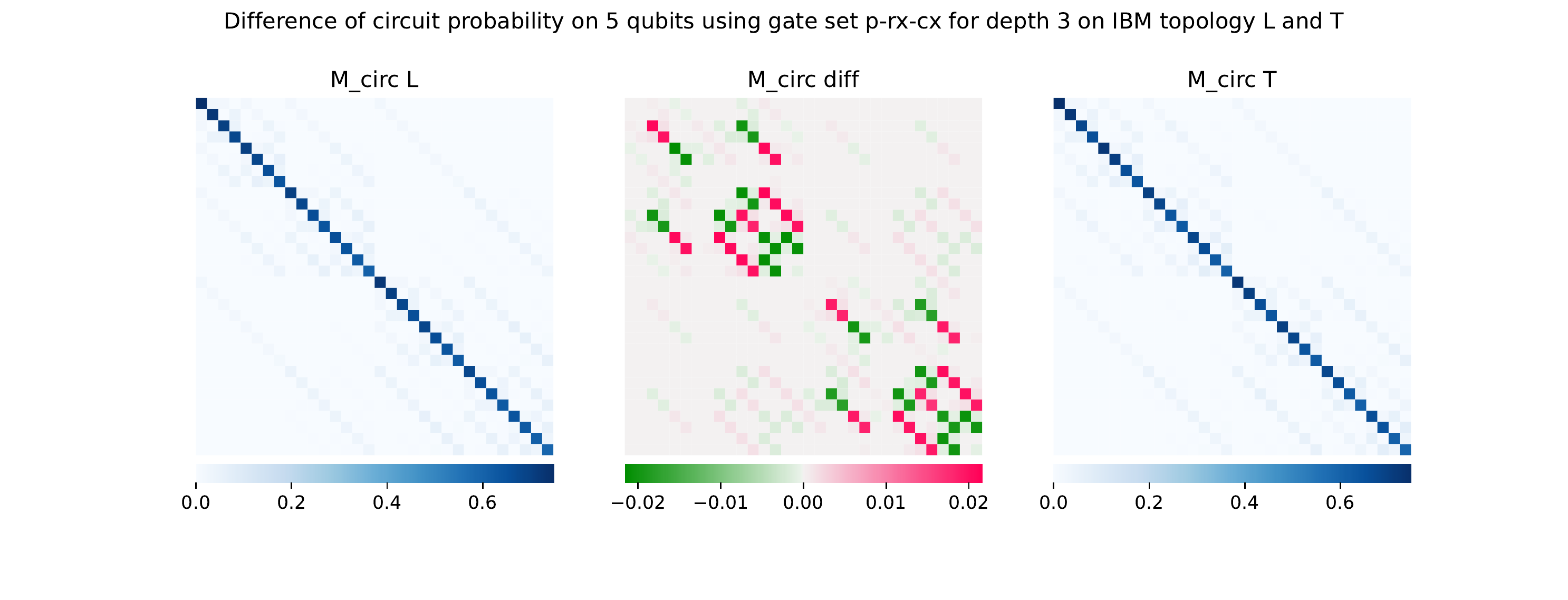}
	\caption{$M_{circ}$ and their comparison for gate set $\mathtt{\{P(\pi/4),RX(\pi/2),CX\}}$ on $5$ qubits and of circuit depth from 0 to 3 on the IBM L-topology and T-topology.}
	\label{fig:ltdiff}
\end{figure}

Using these reachability matrices, we can now calculate the $M_{circ}$ for the two device qubit connectivity topology.
The circuit probability can thereafter be compared with each other.
This gives us insight on which device is better in terms of reachability analysis.
The results, as shown in Figure~\ref{fig:ltdiff}, informs that for some transformations, the L-topology is better and has a higher probability and thereby lower circuit complexity (i.e., the red ones in the middle difference plot), while for others, the T-topology is better (i.e., the green ones for which the $M_{circ}$ for T is higher).

% \todo{show difference in reachibility matrix in appendix}

% \newpage

% \begin{figure}[htb]
% 	\centering %LBRT
% 	\includegraphics[width=0.6\textwidth]{figures/growth_rate_prog.pdf}
% 	\caption{Growth of the number of permutation of gates with qubit count for two types of gate sets: (i) $[1,3]$ qubits: $\mathtt{\{X,CCX\}}$, (ii) $[1,1,2]$ qubits: $\mathtt{\{H,S,CX\}}$, $\mathtt{\{H,T,CX\}}$, $\mathtt{\{P(\pi/4), RX(\pi/2), CX\}}$}
% 	\label{fig:growth}
% \end{figure}

% An example of $M_n$ can be shown below. In the example, with matrix $M_1$, with only two gate X and CCX, there are 16 distinct circuits can be formed. This is reason why the sum of all row is equal to 16. Moreover, if we have a look at element $A_{0, 0}$ it represent the fact that state 0000 to 0000 with 12 possible way (12 possible circuits formed by CCX gate). 
% \begin{verbatim}

% \end{verbatim}
% The degree of the matrix $M_n$ and $\overline{M}_n$ is related to the number of gates.

The Expressibility plots follow a fractal structure, which is effectively the trace of the subsystem, as:

\begin{equation}\label{eq:exp}
E_i^n = \begin{bmatrix}
    A_i^n & B_i^n \\
    B_i^n & A_i^n
    \end{bmatrix} =
    \begin{bmatrix}
    B_{i+1}^n & A_{i-1}^n \\
    A_{i-1}^n & B_{i+1}^n
    \end{bmatrix},\quad
    A_i^n = E_i^{n-1},\quad
    B_i^n = E_{i-1}^{n-1}
\end{equation}

\newpage
    
Another important insight is that the reachability/expressibility analysis is independent of the gate set being strongly universal (e.g. $\mathtt{\{H,~T,~CX\}}$) or not (e.g. $\mathtt{\{H,~S,~CX\}}$).
This is due to our focus on the existence of a path that maps between two states, without considering the how much we can control and steer the system towards a specific path.
To illustrate the point, a gate set of just $\mathtt{\{H\}}$ or $\mathtt{\{X\}}$ can map between any pair of states (weakly universal) given sufficient depth, yet they clearly cannot approximate even classical ones like $\mathtt{\{NAND\}}$, let alone all functions.
To define universality in this framework, we define functions as a set of transformations between states (e.g. as a sum-of-product expression, probability mass function, or an unitary matrix) 
A universal gate set that can approximately represent any such transformation given sufficient depth.
We leave further discussion on universal gate set for our ongoing research as an extension of this work.

\section{Applications}\label{s:application}

The application of this research is primarily twofold.
On one hand, it is an exploration via enumeration of the characteristics of Hilbert space.
A visual map of the structures presented in the results would aid in an intuitive understanding of the capabilities of quantum computation.
This is an extension of similar projects in classical logic~\cite{demey2019metalogic} and algorithmic information~\cite{zenil2021computable}.
On a more pragmatic footing, this research finds applications for various use cases.
% \begin{enumerate}[nolistsep,noitemsep]
%     \item Geometric quantum machine learning
%     \item Novel quantum algorithm synthesis
%     \item Quantum artificial general intelligence
% \end{enumerate}
We conclude this article with a brief description of how this research connects to these use cases.

\subsection{Geometric quantum machine learning}

The landscape of quantum processes is of interest for both foundational and practical aspects of quantum information.
On the foundational side, quantum complexity theory~\cite{brown2018second,haferkamp2022linear}, quantum resource theories~\cite{halpern2022resource}, categorical quantum mechanics~\cite{selby2021reconstructing} and quantum formal logic~\cite{pratt1993linear} rely on the properties of this landscape.
The transition to practical aspects is orchestrated by the geometric formulation of quantum computation~\cite{nielsen2006quantum,nielsen2006optimal}.
Recently, this forms the basis for quantizing geometric deep learning~\cite{perrier2020quantum}.
These works have been conducted in the formalism of mathematical functions or quantum fields\cite{general2018principle}.
Circuit complexity is much less studied, and can bridge algorithmic complexity and computational complexity.
By providing a perspective of the statistical/algorithmic complexity geometry of quantum logic circuits, our intention is to make these results tangible using quantum computational frameworks in the near future.
On the other hand, operational distance measures between two quantum states/processes for specific use cases can be informed by these theoretical techniques.

% quantum supremacy

% (speculative) understand if random circuit sampling is same as quantum algorithmic probability, have an use for that experiment. 
% Use QPULBA for that.
% https://arxiv.org/pdf/2207.14280.pdf random circuits in field theory, quantum chaos, ..

% Lindenbaum–Tarski algebra
% https://en.wikipedia.org/wiki/Lindenbaum%E2%80%93Tarski_algebra

\subsection{Novel quantum algorithm synthesis}

Quantum algorithm design currently involves a careful manipulation of quantum information, harnessing quantum mechanical phenomena (e.g. superposition, entanglement, interference) to a computational advantage.
This is generally counter-intuitive to human phenomenological experiences, thus requiring considerable training and often serendipitous moments~\cite{shor2022early}.
Though discovery of new algorithms is an buzzing research field~\cite{jordan}, reasoning in terms of mathematical formalism has been a barrier to wider adoption of quantum accelerated computing.
Some proposals for automation of quantum programming~\cite{spector2004automatic,quetschlich2022towards,sarkar2022automated} has been proposed to remedy these issues.
To further expand the applicability of quantum algorithms, techniques from novelty search~\cite{lehman2010efficiently} and large language models~\cite{lehman2022evolution} can be incorporated into these automation engines.
Open-ended search in the space of quantum processes can greatly benefit from a characterization of this landscape, as presented in this work.

\subsection{Quantum artificial general intelligence}

% The spectrum spanning geometric quantum machine learning~(GQML) and novel quantum algorithm synthesis~(NQAS) pertains to quantum artificial general intelligence~(QAGI).
Among the more rigorous methods of developing general intelligence is an active formulation of Solomonoff's theory of inductive intelligence~\cite{solomonoff1964formal}, called universal artificial intelligence~\cite{hutter2004universal}.
Universal reinforcement learning models like AIXI and KSA are capable of rational decision-making or modelling environmental dynamics, based on the shortest program that corresponds to compressing the past observations and maximizing a set reward function.
These have been quantized both by using quantum algorithms, (e.g., in the AIXI-q model~\cite{catt2020gentle}) and by applying them to quantum environments (e.g., in the QKSA model~\cite{sarkar2023qksa}).

Another crucial aspect of intelligence~\cite{lavin2021simulation} is the understanding of cause-effect relations.
Quantum acceleration of causal inference~\cite{sarkar2021estimating,chiribella2019quantum,acharya2022quantum} can benefit from the knowledge of the probability distribution of causal oracles, a subset of quantum processes that embed specific properties of the problem.
Besides causal inference, similar techniques can be applied to other statistical relational learning applications like probabilistic logic networks~\cite{wittek2017quantum} and quantum variational algorithms. 
% ~\cite{benedetti2021variational}

Both universal distribution and causal inference are intimately connected to the landscape of quantum programs.
This landscape inturn depends on the choice of a specific gate set, as we saw in this research.
Thereby, novelty seeking in the space of universal gate sets can meta-optimize quantum program synthesis for specific application algorithms.
In our current research, we are exploring this direction of second-order cybernetics of automated quantum operational theory, by using the groundwork developed in this article.

% \section{Conclusion} \label{s:conclusion}

% \textcolor{red}{TBD:}
% \begin{itemize}[nolistsep,noitemsep]
%     \item Summary of main motivation and results
%     \item Future work - realistic connectivity, hardware errors, causality, yaqq, causal cone of ancilla states
% \end{itemize}

% \newpage
\section*{Acknowledgements}

This project was initiated under the QIntern 2021 project ``Reinforcement Learning Agent for Quantum Foundations".
B.G.B., T.A., A.S. would like to thank the organizers of the program and QWorld Association. 
A.K. was partially supported by the Polish National Science Center (NCN) under the grant agreement \verb|2019/33/B/ST6/02011|.
A.S. acknowledges funding from the Dutch Research Council (NWO) through the project ``QuTech Part III Application-based research" (project no. \verb|601.QT.001 Part III-C - NISQ|).

\section*{Author contributions}

Conceptualization, A.S.; 
methodology, A.S., A.K. and B.G.B.; 
software, B.G.B. and A.K.; 
% validation, B.G.B. and A.K.; 
% formal analysis, X.X.; 
% investigation, X.X.; 
% resources, X.X.; 
% data curation, B.G.B. and A.K.; 
writing--original draft preparation, A.S., A.K. and B.G.B.; 
% writing--review and editing, T.A. and A.S.; 
visualization, A.K. and T.A.; 
supervision, A.S.
% project administration, A.S.; 
% funding acquisition, Y.Y. 
\\All authors have read and agreed to the published version of the manuscript.

% Please turn to the  \href{http://img.mdpi.org/data/contributor-role-instruction.pdf}{CRediT taxonomy} for the term explanation. 
% Authorship must be limited to those who have contributed substantially to the work reported.

% \vspace{1em} \hline \vspace{0.5em}

% \textcolor{red}{List of ToDos:}
% \begin{itemize}[nolistsep,noitemsep]
%     \item[\scriptsize$\blacksquare$] Growth rate of number of programs with qubit count (on X-axis) for each gate-set (multi-line plot).
%     \item[\scriptsize$\blacksquare$] Reachability and expressibility plot for increasing operation and qubits for each gate-set. Max 4 or 5 qubits, Max 4 or 5 operations (decide based on how large the set is). (Max 4 depth, 4 qubits)
%     \item[\scriptsize$\blacksquare$] Comparative study of universal vs. non-universal
%     \item[\scriptsize$\blacksquare$] Comparative study of quantum vs. classical
%     \item[\scriptsize$\blacksquare$] $M_n = (M_1)^n$ 1-step is enough for n-hop paths; reachability to all binary states; AP 
%     \item[\scriptsize$\blacksquare$] AP idea works for other basis and quantum?
%     \item[\scriptsize$\blacksquare$] Plot M circ for x-ccx and h-t-cx 
%     \item[\scriptsize$\blacksquare$] Expressibility fractal nature
%     \item[\scriptsize$\blacksquare$] Variation of expressibility/reachability for IBM gate-sets with 5 qubits for all-to-all connection, 5-qubit IBM T-shaped topology, 5-qubit line topology.
%     \item[$\square$] Noise simulation
% \end{itemize}

% \vspace{0.5em} \hline \vspace{1em}

% \newpage
\bibliographystyle{unsrt}
{\footnotesize
\bibliography{ref.bib}}

% \appendix
% \input{appendix}

\end{document}